\documentclass[conference, nofonttune, 9pt]{IEEEtran}
\pdfoutput=1
\usepackage{lmodern}
\usepackage{amsmath} %
\usepackage{amssymb} %
\usepackage{graphicx}
\usepackage{hyperref}
\usepackage{multirow}
\usepackage{xcolor}
\usepackage{subcaption}
\usepackage{makecell}
\usepackage{enumitem}
\usepackage{fancyhdr}
\usepackage{booktabs}

\usepackage[bibstyle=ieee, citestyle=numeric-comp, sortcites=true, backend=bibtex, maxbibnames=99]{biblatex}
\IEEEoverridecommandlockouts

\usepackage{tikz,pgfplots}

\hypersetup{
    colorlinks=blue!80!black,
    pdfborder={0 0 0},
}

\usepackage[protrusion, expansion]{microtype}

\pgfplotscreateplotcyclelist{MarkerCycleList}{%
    solid, 			every mark/.append style={solid, fill=white}, mark=*\\%
    dashed, 		every mark/.append style={solid, fill=white}, mark=diamond*\\%
    dashed, 		every mark/.append style={solid, fill=white}, mark=triangle\\%
    densely dotted, every mark/.append style={solid, fill=white}, mark=star\\%
    solid,          every mark/.append style={solid, fill=white}, mark=Mercedes star flipped\\%
    densely dashed, every mark/.append style={solid, fill=white}, mark=otimes*\\%
}

\pgfplotsset{compat=1.3}

\begin{document}

\title{A High-Quality Workflow for Multi-Resolution Scientific Data Reduction and Visualization}

\author{\normalsize\null\\[-4.2ex] %
Daoce Wang\IEEEauthorrefmark{1},
Pascal Grosset\IEEEauthorrefmark{2},
Jesus Pulido\IEEEauthorrefmark{2},
Tushar M. Athawale\IEEEauthorrefmark{3},
Jiannan Tian \IEEEauthorrefmark{1},
Kai Zhao \IEEEauthorrefmark{4},\\
Zarija Lukić\IEEEauthorrefmark{5},
Axel Huebl\IEEEauthorrefmark{5}, 
Zhe Wang\IEEEauthorrefmark{3},
James Ahrens\IEEEauthorrefmark{2},
Dingwen Tao\IEEEauthorrefmark{6}\thanks{Dingwen Tao is the corresponding author.} \thanks{Kai Zhao and Jiannan Tian's participation in this work ended on 03/26/2024.}\\[.9ex]  %
\IEEEauthorrefmark{1} Indiana University, Bloomington, IN, USA;\ \  \texttt{\string{\,daocwang,\,jti1\,\string}@iu.edu}\\
\IEEEauthorrefmark{2}
Los Alamos National Laboratory, Los Alamos, NM, USA;\ \  \texttt{\string{\,pascalgrosset, pulido, ahrens\,\string}@lanl.gov} \\
\IEEEauthorrefmark{3}
Oak Ridge National Laboratory, Oak Ridge, TN, USA;\ \  \texttt{\string{\,athawaletm, wangz\,\string}@ornl.gov}\\
\IEEEauthorrefmark{4}
Florida State University, Tallahassee, FL, USA;\ \  \texttt{kai.zhao@fsu.edu}\\
\IEEEauthorrefmark{5}
Lawrence Berkeley National Laboratory, Berkeley, CA, USA;\ \  \texttt{\string{\,zarija, axelhuebl\,\string}@lbl.gov}\\
\IEEEauthorrefmark{6}
SKLP, Institute of Computing Technology, Chinese Academy of Sciences, China;\ \ \texttt{taodingwen@ict.ac.cn}\\[-3ex]  %
}

\maketitle

\thispagestyle{plain}
\fancypagestyle{plain}{%
  \renewcommand{\headrulewidth}{0pt}%
  \fancyhf{}%
  \fancyfoot[L]{\footnotesize SC24, November 17-22, 2024, Atlanta, Georgia, USA 
  \newline 979-8-3503-5291-7/24/\$31.00 \copyright 2024 IEEE}%
}

\begin{abstract}

Multi-resolution methods such as Adaptive Mesh Refinement (AMR) can enhance storage efficiency for HPC applications generating vast volumes of data. However, their applicability is limited and cannot be universally deployed across all applications. Furthermore, integrating lossy compression with multi-resolution techniques to further boost storage efficiency encounters significant barriers. To this end, we introduce an innovative workflow that facilitates high-quality multi-resolution data compression for both uniform and AMR simulations. Initially, to extend the usability of multi-resolution techniques, our workflow employs a compression-oriented Region of Interest (ROI) extraction method, transforming uniform data into a multi-resolution format. Subsequently, to bridge the gap between multi-resolution techniques and lossy compressors, we optimize three distinct compressors, ensuring their optimal performance on multi-resolution data. These optimizations can improve the compression ratio of SOTA approaches by up to 3.3$\times$ under the same data quality loss. Lastly, we incorporate an advanced uncertainty visualization method into our workflow to understand the potential impacts of lossy compression. Experimental evaluation demonstrates that our workflow achieves significant compression quality improvements.

\end{abstract}

\IEEEpeerreviewmaketitle

\section{Introduction}

In recent years, the complexity and costs associated with scientific simulations have significantly increased. To address these challenges, numerous HPC simulation tools have adopted multi-resolution methods, such as the Adaptive Mesh Refinement (AMR) technique~\cite{zhang2019amrex, stone2020athena++, Flash-X-SoftwareX}. AMR aims to reduce computational expenses while preserving the accuracy of simulation outcomes. Unlike traditional uniform mesh techniques that apply consistent resolution throughout the simulation space, AMR employs a dynamic approach. It selectively increases resolution in regions of interest, thereby optimizing computational resource usage and minimizing storage requirements.

While AMR offers significant benefits in terms of computational, storage, and memory efficiency, its implementation in some scientific simulations is hindered by several challenges.
First, integrating AMR can be technically demanding, requiring substantial modifications to existing numerical algorithms and simulation codes, which may not be feasible for all projects. In some instances, simulation algorithms might not accommodate specific geometries or the dynamic adjustments of the grid throughout the simulation's evolution.
Additionally, AMR algorithms introduce complexity in grid management and error control, posing optimization challenges for certain simulations, especially those involving highly complex phenomena like convex plasma shapes.
For example, in WarpX electromagnetic simulation~\cite{warpx}, mesh refinement is currently restricted to disjoint cuboids, which limits the full flexibility offered by AMR~\cite{VayPML2004,Vay2013}. %

In order to enable uniform-grid simulations to benefit from multi-resolution storage, thereby reducing disk usage, I/O (input/output) time, and memory footprint in visualization without complicating the simulation process, previous work~\cite{kumar2014efficient, tian2013dynam, bhatia2022amm, usher2021adaptive} has adapted the multi-resolution storage approach for uniform grids. These methods can, for example, store regions of interest at full resolution while representing less critical areas at a lower resolution for visualization or analysis. \textit{However, the space saved from using multi-resolution alone is often not enough.}
For instance, a multi-resolution dataset with $0.5 \times 1024^3$ mesh points at the coarse level and $0.5 \times 2048^3$ at the fine level could yield about 1 TB of data per snapshot.
Consequently, conducting five simulations with 200 snapshots would require a total disk storage of 1 PB.
Simulations used in Exascale scenarios can be even larger than that, using many thousands of points per axis~\cite{warpx}, making data size reduction a timely need.

To this end, data compression can be utilized alongside multi-resolution techniques to further reduce I/O and storage costs. However, traditional lossless compression methods provide limited data volume reduction for scientific simulations, typically achieving compression ratios of only up to 2$\times$.
As a solution, a new generation of error-bounded lossy compression techniques, such as SZ~\cite{tao2017significantly, di2016fast, sz18}, ZFP~\cite{zfp}, MGARD \cite{ainsworth2018multilevel} and their GPU versions \cite{tian2020cusz,tian2021optimizing, cuZFP}, have been widely used in the scientific community ~\cite{jin2024concealing,Huebl2017,zfp,sz18,lu2018understanding,luo2019identifying,tao2019optimizing,cappello2019use,jin2020understanding,grosset2020foresight,jin2022accelerating, baker2019evaluating,di2024survey} due to their ability to offer high compression ratios while maintaining controllable accuracy impacts on various scientific applications.

While lossy compression has the potential to significantly reduce I/O and storage costs for multi-resolution data, its effective application in this context remains under-explored.
Three recent studies have targeted the development of efficient lossy compression methods for multi-resolution data including AMR data. zMesh~\cite{zMesh} was proposed to reorder AMR data using z-order across different refinement levels into a 1D array, leveraging data redundancy. However, zMesh cannot leverage higher-dimension compression by compressing data in a 1D, leading to a loss of spatial information in higher-dimension data. On the other hand, TAC~\cite{wang2022tac,wang2024tac+} improved zMesh's compression quality through adaptive 3D compression. While zMesh and TAC offer offline compression solutions for AMR data, they did not delve into in-situ compression, which could notably reduce the I/O cost. AMRIC~\cite{amric} addressed this by introducing an in-situ AMR compression framework designed to lower I/O costs while improving compression quality for AMR applications.

These efforts have primarily focused on optimizing multi-resolution data compression for block-wise compressors like SZ2 and ZFP. The block-wise nature of these compressors enables higher speed but also renders them susceptible to compression artifacts due to the loss of spatial information between blocks. In contrast, non-block-wise (global) compressors like SZ3, despite their lower throughput, often achieve better compression quality in most cases by leveraging prediction across the entire input data. However, SZ3's compression approach presents significant challenges when applied to multi-resolution data, a topic that will be further discussed in \S\ref{sec:pad}.

To this end, this paper proposes a comprehensive workflow for compressing multi-resolution data, suitable for both adaptive data derived from uniform-resolution simulations and AMR data. Our strategy not only addresses SZ3's performance issues with multi-resolution data but also introduces a novel post-processing technique to enhance data quality from block-wise compressors like SZ2 and ZFP. 
\textcolor{black}{
Moreover, compression may result in compression artifacts, and there has been no study on identifying potential compression artifacts.} Thus, we explore the ramifications of compression-induced uncertainty, aiding users in understanding how compression affects their data.

Our primary contributions are as follows:
\begin{itemize}[leftmargin=1.3em]

    \item We employ a compression-oriented, adaptive region-of-interest (ROI) method to convert uniform data into multi-resolution data, thereby enhancing storage efficiency while maintaining the quality of visualization and post-analysis.
    \item We propose SZ3MR, an optimization of the state-of-the-art global lossy compressor SZ3 for multi-resolution data, incorporating dynamic padding and adaptive error bounds within the SZ3 compressor to improve prediction accuracy and compression quality.
    \item We develop an efficient and effective error-bounded post-processing solution that leverages spatial information across each compressed block to significantly enhance the quality of block-wise compressors (e.g., SZ2/ZFP). This solution is also adaptable to improving multi-resolution data compression with global compressors like SZ3.
    \item  We investigate the uncertainty introduced by lossy compression, an under-explored topic, by integrating a cutting-edge uncertainty visualization technique. This enables a clearer understanding of how compression affects the data through visual representation (will be detailed in \S\ref{sec:unc}).
    \item Our experiments show significant compression performance improvements with low overhead for five scientific applications. Our workflow is also integrated into real-world scientific applications, WarpX and Nyx, for in-situ processing.
\end{itemize}

\section{Background}
\subsection{Lossy Compression for Scientific Data}

Recent research has introduced high-precision lossy compression algorithms for scientific data, notably SZ~\cite{sz17, sz18, zhao2021optimizing}, ZFP~\cite{zfp}, MGARD~\cite{gong2023mgard}, and TTHRESH~\cite{ballester2019tthresh}, which differ from traditional compressors like JPEG by targeting floating-point data with strict error control based on user requirements.
This work focuses on three compression algorithms: SZ2, ZFP, and SZ3. The key difference between them is that SZ2 and ZFP are block-wise, while SZ3 is global (non-block-wise). SZ2 and ZFP partition the input data into smaller blocks (e.g., $4\times4\times4$ for ZFP) and process them separately to leverage the spatial information.
Specifically, SZ2 uses the Lorenzo predictor or linear regression for each block, and ZFP applies a DCT-like transform.
In contrast, SZ3 employs global interpolation prediction across the entire input data without partitioning it.
SZ2 and ZFP offer fast compression speeds, but the global interpolation of SZ3 enables it to capture more spatial information across the dataset, thus producing a higher compression quality/ratio than the block-wise SZ2/ZFP. We refer readers to~\cite{sz18, zfp, zhao2021optimizing} for more details.

\subsection{AMR Method and Multi-resolution Data}
\begin{figure}[h]
    \vspace{-2mm}
    \centering
        \includegraphics[width=.63\linewidth]{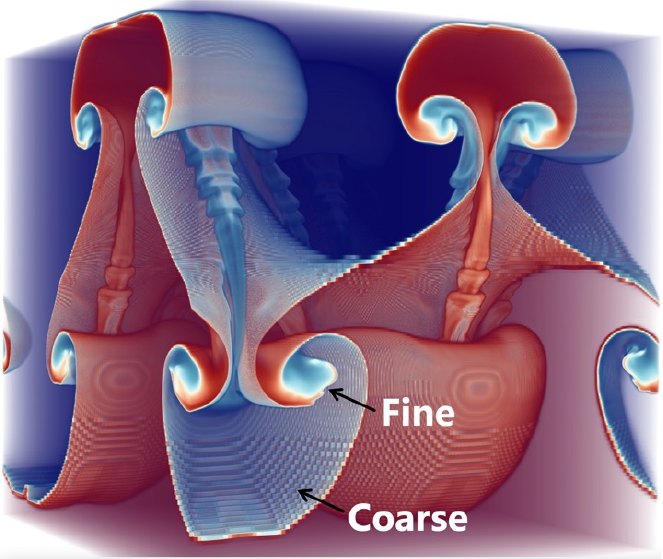}
        \caption{Example of an AMR dataset of Rayleigh–Taylor instability.}
        \label{rt-vis}
    \vspace{-3mm}
\end{figure}

By using a non-uniform grid, AMR can significantly enhance computational efficiency and lower storage requirements while still achieving the desired accuracy level.
In AMR applications, the mesh or spatial resolution is dynamically adjusted according to the simulation's demands, implementing a finer mesh in areas of greater significance or interest and a coarser mesh in less critical regions as depicted in Fig.~\ref{rt-vis}. In AMR application, the mesh is refined based on specific criteria, such as when the average value of a block exceeds predefined thresholds.

\begin{figure}[h]
    \vspace{-1mm}
    \centering
    \begin{subfigure}[t]{0.28\linewidth}
        \centering
        \includegraphics[width=\linewidth]{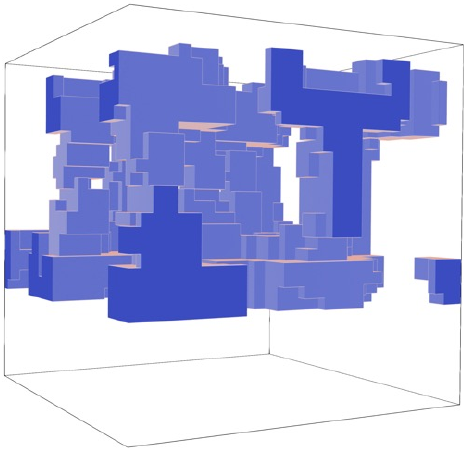}
        \caption[t]{Fine level}
        \label{fig:dis_fine}
    \end{subfigure}
    \begin{subfigure}[t]{0.28\linewidth}
        \centering
        \includegraphics[width=\linewidth]{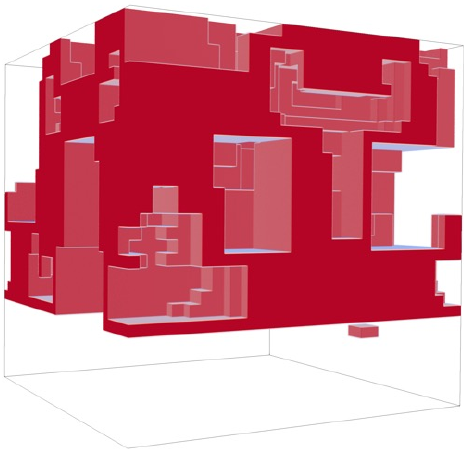}
        \caption{Mid level}
        \label{fig:dis_mid}
    \end{subfigure}
    \begin{subfigure}[t]{0.28\linewidth}
        \centering
        \includegraphics[width=\linewidth]{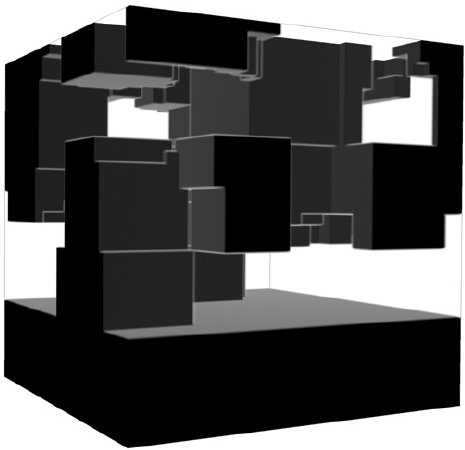}
        \caption{Coarse level}
        \label{fig:dis_coarse}
    \end{subfigure}
    \caption[t]{Vis of data distributions for different  level for Fig.~\ref{rt-vis}.}
    \vspace{-1mm}
    \label{fig:dis}
\end{figure}

For non-AMR (uniform) simulation, one can also achieve storage efficiency by storing important regions at full resolution and nonessential regions at lower resolution. For example, Previous work~\cite{kumar2014efficient} proposes using range thresholding to identify ROIs and reduce non-ROI resolution and then using HZ ordering to traverse all the resolution levels to benefit the I/O. 
However, HZ-ordering prevents us from achieving optimal compression performance because it flattens high-dimensional data into 1D, resulting in the loss of spatial information. At the same time, many previous studies~\cite{sz17,zhao2020significantly,wang2022tac} proved that leveraging more spatial information can significantly improve compression performance. To compress the data in 3D, we propose compression-oriented importance-driven storage of uniform data by processing different resolution levels separately (see \S\ref{sec:de}).

The multi-resolution data, including AMR data and the adaptive data generated from uniform data, are hierarchical with different resolutions, with each resolution level holding a different part of the domain, as illustrated in Fig.~\ref{fig:dis}.

\subsection{Uncertain Data and Visualization}

Significant research has been conducted on effective methods for visualizing uncertain scientific data~\cite{TA:Johnson:2003:nextStepVisErrors, TA:2016:Ferstl:isosurfaceUncertainty, TA:Athawale:2021:nonparametricDVR,TA:Otto:2011:3dVectorFieldTopologyUncertainty, TA:Wang:2019:uncertaintyEnsemble, jiao:TensorUncertainty:2012}, as not knowing uncertainty in data can lead to incorrect scientific conclusions. Uncertainty in data often arises from inaccuracies in data acquisition or due to the limitations and incompleteness of measurements available for computational simulations~\cite{ TA:Brodlie:2012:RUDV}. Similarly, uncertainties in model parameters during scientific simulations introduce variability into the computed solutions~\cite{TA:Potter:2012:UQtaxonomy}.
Uncertain data is typically represented by probability distributions at each data point~\cite{DavidJosephJulien2014,VietinghoffBaysianCriticalPointUncertainty, TA:Athawale:2021:topoMappingUncertaintyMarchingCubes}, in contrast to deterministic data, which assigns a specific value to each point.

The compression techniques, when applied to the original data, can result in a loss of information and introduce error/uncertainty in decompressed data.
However, there has been a gap in research regarding treating decompressed data as a form of uncertain data and using uncertainty visualization techniques to explore the effects of compression on scientific datasets.
In our work, we apply cutting-edge uncertainty visualization techniques to decompressed data, aiming to provide a clearer understanding of the potential impacts of the compression (see \S\ref{sec:unc}).

\section{Our Proposed Design}
\begin{figure}
  \centering
  \includegraphics[width=\linewidth]{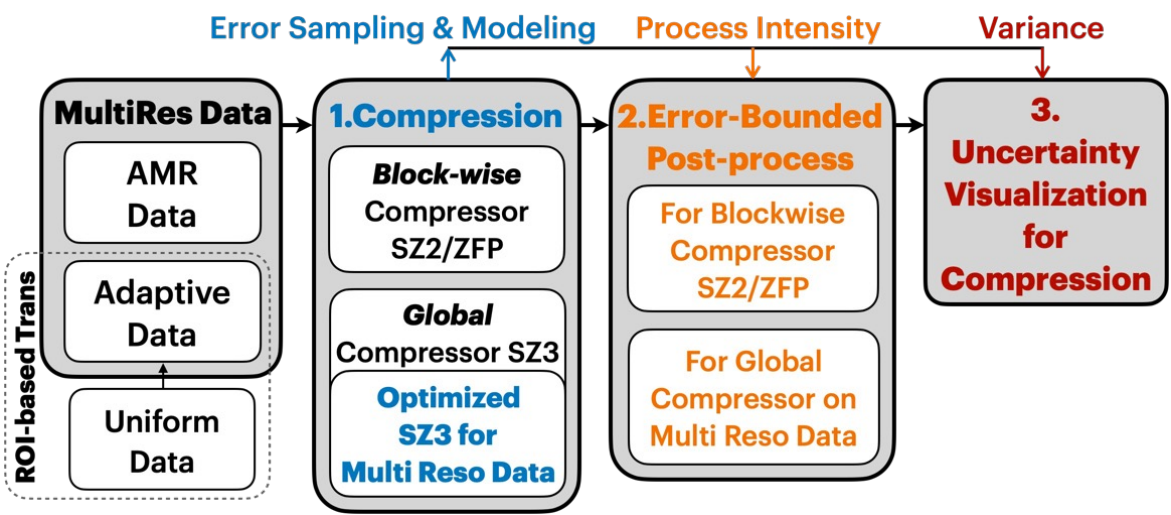}
  \caption{Overview of our proposed workflow for multi-resolution scientific data compression.}
  \label{fig:overview}
\end{figure}

This section outlines our proposed workflow for multi-resolution data compression, as shown in Fig.~\ref{fig:overview}.
In \S\ref{sec:pad},  We detail our optimization of the SZ3 compressor for multi-resolution data compression (SZ3MR). By employing dynamic padding and an adaptive error-bound approach considering the features of multi-resolution data,
we significantly improve SZ3's compression performance on multi-resolution data.

In \S\ref{sec:post}, we improve the decompressed data quality from block-wise compressors (e.g., SZ2 and ZFP). We introduce a dynamic, error-bounded post-processing technique that optimally utilizes the spatial information within the dataset. This versatile post-processing method can also improve multi-resolution data compression when using global compressors like SZ3.

In \S\ref{sec:unc}, we explore the uncertainties introduced by the compression. By integrating a cutting-edge uncertainty visualization solution, we provide users with insights into how compression may impact the data. This exploration aids in understanding and mitigating the effects of compression error.

\label{sec:de}
\begin{figure}
  \centering
  \begin{subfigure}[t]{0.47\linewidth}
    \centering
    \includegraphics[width=\linewidth]{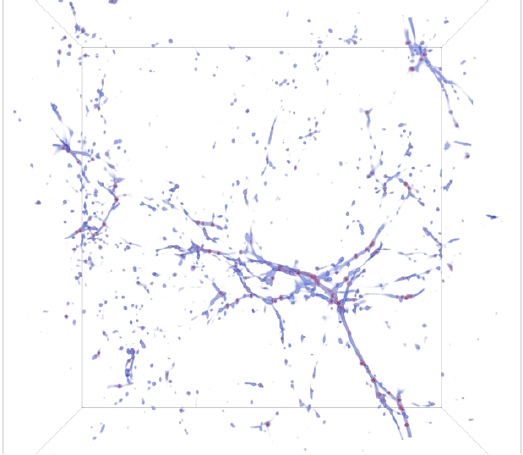}
    \caption[t]{Original data}
    \label{fig:roi-all}
  \end{subfigure}
  \begin{subfigure}[t]{0.47\linewidth}
    \centering
    \includegraphics[width=\linewidth]{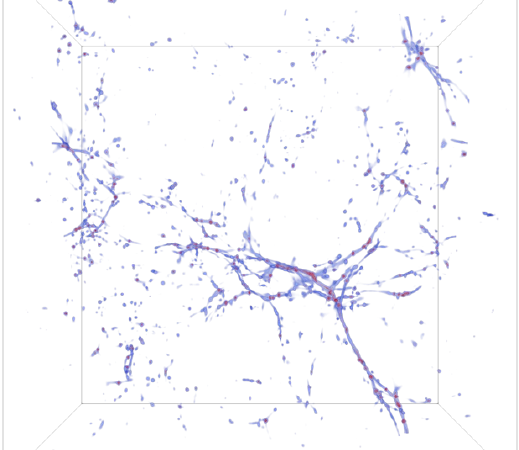}
    \caption{ROI}
    \label{fig:roi-fine}
  \end{subfigure}

  \caption[t]{Visualization of the original Nyx cosmology dataset (left) and the ROI (right, 15\% of the dataset) extracted using our approach, the SSIM of the two pictures is 0.99995.}
  \label{fig:roi}
\end{figure}

\begin{figure*}[t]
  \centering
  \begin{subfigure}[t]{0.25\linewidth}
    \centering
    \includegraphics[width=\linewidth]{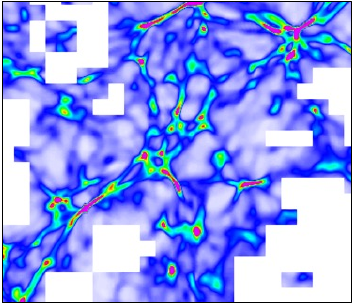}
    \caption[t]{Original data}
    \label{fig:nyx-slides-ori}
  \end{subfigure}~%
  \begin{subfigure}[t]{0.25\linewidth}
    \centering
    \includegraphics[width=\linewidth]{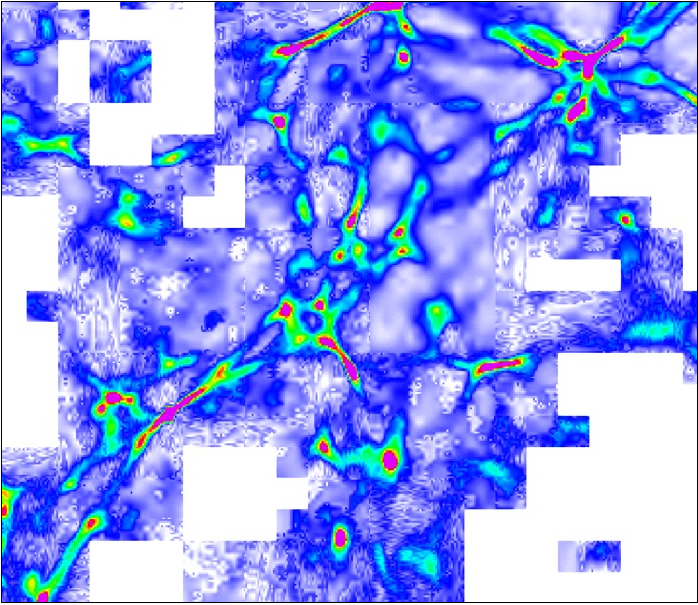}
    \caption{TAC, SSIM=.64, PSNR=117.6}
    \label{fig:nyx-slides-tac}
  \end{subfigure}~%
  \begin{subfigure}[t]{0.25\linewidth}
    \centering
    \includegraphics[width=\linewidth]{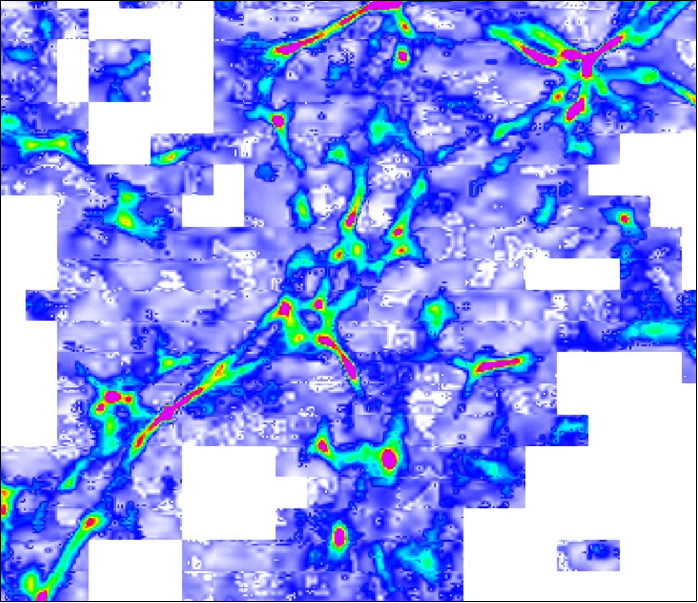}
    \caption{AMRIC, SSIM=.57, PSNR=115.0}
    \label{fig:nyx-slides-amric}
  \end{subfigure}~%
  \begin{subfigure}[t]{0.25\linewidth}
    \centering
    \includegraphics[width=\linewidth]{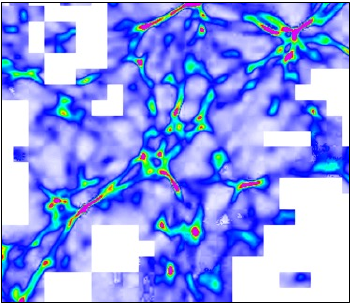}
    \caption{Ours, SSIM=.91, PSNR=123.4}
    \label{fig:nyx-slides-ours}
  \end{subfigure}
  \caption[t]{Vis comparison (one $1.5\times$ zoom in 2D slice) of original data and decompressed data produced by TAC's SZ3, AMRIC's SZ3 and our SZ3MR on Nyx's ``baryon density'' field (fine level). Warmer colors indicate higher values. The CR of TAC, AMRIC, and ours is the same, 163.}
    \vspace{-5mm}
  \label{fig:nyx-slides}
\end{figure*}

\textbf{\textit{ROI selection and preprocessing of multi-resolution data.}} We will first introduce how we convert uniform data into multi-resolution data (referred to as adaptive data) and then detail the preparation of the multi-resolution data (including adaptive data and AMR data) for 3D compression.

We begin by partitioning the original dataset into blocks of size $b\times b\times b$, where $b$ is $2^n, (n > 2)$. Then, following the method of \cite{kumar2014efficient}, we utilize range thresholding to identify ROIs due to its lightweight and effective characteristics. Specifically, we calculate each block's value range and select the top $x$ percent of the blocks as the ROIs ($x=50\%$ by default, adjustable for specific applications). Non-ROI blocks are stored at a lower resolution to enhance storage efficiency. For example, as shown in Fig.~\ref{fig:roi}, our range-based ROI selection method effectively extracts the over-density halos from the Nyx cosmology dataset. By selecting just 15\% of the dataset, we can capture almost all the halos for the Halo-finder analysis of Nyx~\cite{davis1985evolution}.

\begin{figure}
  \centering
  \includegraphics[width=0.95\columnwidth]{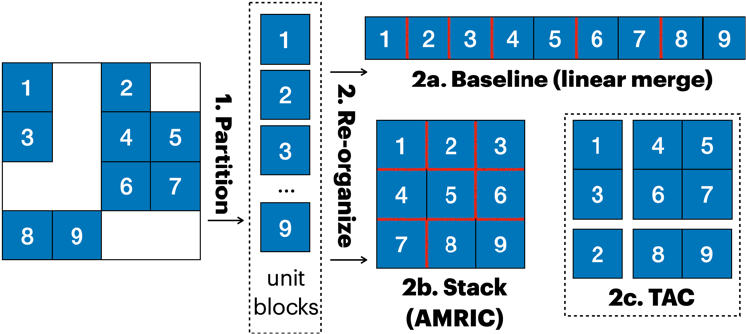}
  \caption{2D Example of uniform partition (left, part 1) and different arrangements (the linear merge baseline, stack merge, and TAC) of the unit block (right, part 2). The bold red line indicates unsmooth boundaries because of the merge of non-neighboring blocks.}
  \label{stack-example}
\end{figure}

After processing, the adaptive data acquires a data structure similar to AMR data. To compress them in 3D, we diverge from the HZ-ordering method used in \cite{kumar2014efficient}, which flattens the data to 1D. Instead, we propose compressing each resolution level separately in 3D. However, as illustrated in Fig.~\ref{fig:dis}, each level exhibits many empty regions and an irregular data distribution. To address this, we employ a uniform partitions method, which divides the data into a collection of 3D ``unit blocks'', as shown in the left part of Fig.~\ref{stack-example}, for later process and compression.

\subsection{Improved SZ3 for multi-resolution data (SZ3MR)}
\label{sec:pad}
The processed multi-resolution data, however, faces a significant challenge that prevents it from achieving optimized compression quality with the original SZ3 compressor. To overcome these challenges and enhance compression performance, we propose optimizing SZ3. As shown in Fig.~\ref{fig:nyx-slides}, our approach achieves much better data quality than SOTA AMRIC~\cite{amric} and TAC~\cite{wang2022tac} using SZ3.
We will now detail the challenge, the limitations of the current approach, and describe our solution.

\textbf{\textit{Challenge: limit compression performance for SZ3 on multi-resolution data.}}
Previous studies have successfully adapted block-wise compressor SZ2/ZFP to handle AMR data. However, optimizing SZ3 for multi-resolution data introduces considerable challenges. A primary concern with SZ3 is the data needs to be partitioned into small unit blocks to leverage 3D compression as mentioned.
This disrupts the spatial information's integrity and diminishes data smoothness.
When SZ3 confronts partitioned unit blocks from multi-resolution data, the disrupted spatial information can significantly undermine the effectiveness of the interpolation prediction without suitable preprocessing steps.

\textbf{\textit{Limitations of the current solutions. }}
Segmented unit blocks of multi-resolution data can be intuitively linearized (e.g., along $z$-axis) into a large 3D array before the compression
as shown in Fig.~\ref{stack-example}-\textbf{2a}. However, this method can significantly affect the effectiveness of SZ3's interpolation, since the other two dimensions of the merged array (e.g., $x$ and $y$) are small, compromising prediction accuracy (will be detailed later).

A previous study AMRIC~\cite{amric}  presented an alternative approach by arranging unit blocks into a cubic, instead of linear merging. This method aims to enable more balanced prediction for each dimension. However, stacking unit blocks into cubic forms aggregates blocks that are not adjacent in the original dataset. This leads to rapid changes in data values between these non-neighboring blocks, resulting in misprediction and adversely affecting the precision of SZ3's prediction. As depicted in the bottom mid part of Fig.~\ref{stack-example}-\textbf{2b}, the stacking process introduces more unsmoothness to the data than linear merging does (indicated by the bold red line).

Another work TAC~\cite{wang2022tac} adopts a dynamic strategy, such as using a kD tree, to merge more adjacent unit blocks from the original dataset, aiming to enhance data smoothness and locality. This approach is depicted in the bottom right part of Fig.~\ref{stack-example}-\textbf{2c}. However, TAC does not have an in-situ solution because TAC's preprocessing requires reconstructing the entire physical domain's hierarchy, a complex task that incurs high overhead for in-situ data compression.
Also, the challenge of small blocks persists (e.g., block 2 remains small) due to the inherent sparsity of multi-resolution data.
Moreover, because the merged blocks vary in shape, TAC must compress the merged blocks with different shapes separately, which brings encoding overhead.

\textit{\textbf{Improvement 1: Better prediction via Padding.}}
We propose to still linearize unit blocks like baseline to avoid the issue of AMRIC and TAC.
However, compared to the baseline, we introduce a padding strategy aimed at enhancing SZ3's prediction accuracy and compression performance for small unit blocks. This strategy is specifically designed to improve prediction performance for the two smaller dimensions of the large linearized array.
To demonstrate the process and limitations of SZ3 interpolation for small blocks, we present an example using 1D linear interpolation. Although simplified, this example embodies the core principles applicable to more complex scenarios like cubic and 3D interpolation.

Consider a dataset in one dimension containing \(N\) elements. SZ3's interpolation approach happened level by level and begins by predicting the first data point (\(d_1\)) using an initial value of 0 for level 0. Then, for level 1, \(d_1\) is used to predict the final data point (\(d_{N}\)). The interpolation process then proceeds in steps size $S$ of \(2^n\), satisfying the condition:
\[
  2^n < N-1, \quad n \in \mathbb{N}
\]
with \(n\) decreasing each level. Each \(S^{th}\) point, not yet predicted, is interpolated from adjacent steps (e.g., predict \(d_{s+1}\) using \(d_1\) and \(d_{2s+1}\)). Points outside the interpolation range are handled through extrapolation.

\begin{figure}[h]
  \centering
  \includegraphics[width=0.83\columnwidth]{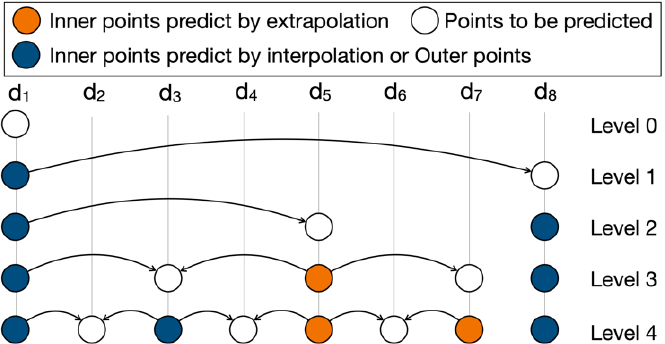}
  \caption{Interpolation example of 8 data points.}
  \vspace{-3mm}
  \label{bad-exp}
\end{figure}

\begin{figure}[h]
  \centering
  \includegraphics[width=0.92\columnwidth]{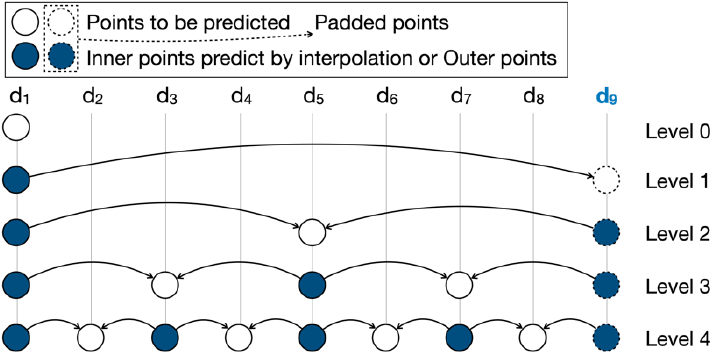}
  \caption{Interpolation example of 9 data points with one padded point.}
  \label{good-exp}
  \vspace{-1mm}
\end{figure}

For small unit blocks partitioned from multi-resolution data, typically of size $2^n$, we examine a scenario with a block size of 8, as shown in Fig.~\ref{bad-exp}. Initially, 0 is used to predict $d_1$, and $d_1$ is used to predict $d_8$ for levels 0 and 1. At level 2, with an interpolation step size of 4, we aim to use $d_1$ and $d_9$ to interpolate $d_5$. However, $d_9$ does not exist, forcing us to depend solely on $d_1$ to extrapolate $d_5$, resulting in limited accuracy.
Similarly, at level 3, only $d_5$ is available for extrapolating $d_7$. After completing the interpolation, it is clear that except for the outer values $d_1$ and $d_8$, 2 out of 6 inner points undergo undesired extrapolation (highlighted in orange). If the block size is 16, this sub-optimal prediction affects 3 out of 14 inner points. Since points predicted at earlier levels are used to predict other points at subsequent levels, the inaccuracies significantly compromise overall compression performance.

To address this issue, we propose the application of padding to the two smaller dimensions of the merged array to enhance prediction performance and eliminate sub-optimal predictions. Given that the multi-resolution data typically adhere to a block size of $2^n$, padding merely requires a single layer of data points to each of the two smaller dimensions (i.e., padding one point for the 1D array), thus introducing acceptable data size overhead. As demonstrated in Fig.~\ref{good-exp}, for a block size of 8, padding an additional point $d_9$ effectively eliminates all sub-optimal predictions for inner data points.
It is also important to determine the pad value, we test using constant, linear, and quadratic extrapolation. After many experiments, we find that the linear extrapolation
overall produce the best prediction performance, especially for the relatively smooth dataset.

On the other hand, while padding can enhance prediction accuracy, it also incurs a size overhead to the data. This overhead is quantified by \((u+1)^2/u^2\), where \(u\) denotes the unit block size. With \(u = 4\), the overhead amounts to 56\%.  In this scenario, padding improves the prediction for 2 out of 3 inner points, but the overall performance gain remains constrained. Moreover, the increased dataset size introduces additional time overhead for compression. Consequently, we opt to implement the padding approach only when \(u > 4\).

As illustrated in Fig.~\ref{fig:Nyx-T2-RT-pad} in \S\ref{sec:eva-off}, our padding strategy, denoted by the curve labeled ``\textbf{\textit{Ours (pad)}}'', significantly enhances the rate-distortion trade-off (PSNR vs. compression ratio) relative to both the AMRIC and baseline. Moreover, our method outperforms the offline-only solution TAC, especially at higher compression ratios.

\textit{\textbf{Improvement 2: Use of adaptive error-bound.}}
We further improve SZ3's performance on multi-resolution data by employing an adaptive error-bound for each interpolation level. This method accounts for the fact that data points predicted at earlier levels influence subsequent-level predictions. For example, as illustrated in Fig.~\ref{good-exp}, point $d_9$ is used for predicting $d_5$, $d_7$, and $d_8$, highlighting the need for smaller error bounds at early interpolation levels to boost compression efficiency.

Although the original SZ3 offers an adaptive error-bound strategy, its coarse granularity limits optimization. Inspired by the QoZ approach~\cite{qoz}, we implemented a more refined adaptive error-bound strategy for each interpolation level:
\[
  eb_l = eb \cdot \left(\min(\alpha^{maxlevel-l}, \beta) \right)^{-1}.
\]
Unlike QoZ, which uses sampling and trial-and-error to select $\alpha$ and $\beta$—a process that introduces overhead—we leverage the characteristics of multiresolution data for a more assertive strategy, setting $\alpha$ to 2.25 and $\beta$ to 8. These parameters are larger than those used by QoZ. This method accelerates the reduction of error bounds for early interpolation levels, particularly for data shapes resulting from linear merges and padding, typically featuring two smaller and one larger dimension (e.g., 17$\times$17$\times$8192). The total interpolation level is low for the two small dimensions, necessitating higher $\alpha$ and $\beta$ to attain small enough error bounds for the initial interpolation levels.
Extensive offline experiment shows that $\alpha=2.25$ and $\beta=8$  deliver the best compression performance in most scenarios.

As illustrated in Fig.~\ref{fig:Nyx-T2-RT-pad} in \S\ref{sec:eva-off}, our approach with padding and adaptive error bound (denoted by the curve ``\textbf{\textit{Ours (pad+eb)}}'') can further improve the compression performance. And, as shown in Fig.~\ref{fig:nyx-slides}, after the two-step optimization, our approach notably improves the overall compression and visualization quality in comparison to the AMRIC and TAC.

\subsection{Error bounded Adaptive post processing}
\label{sec:post}
\begin{figure*}
  \centering
  \begin{subfigure}[t]{0.32\linewidth}
    \centering
    \includegraphics[width=\linewidth]{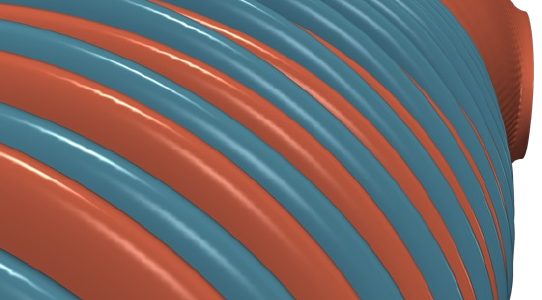}
    \caption[t]{Original data, WarpX}
    \label{fig:post-iso-ori-2}
  \end{subfigure}
  \begin{subfigure}[t]{0.32\linewidth}
    \centering
    \includegraphics[width=\linewidth]{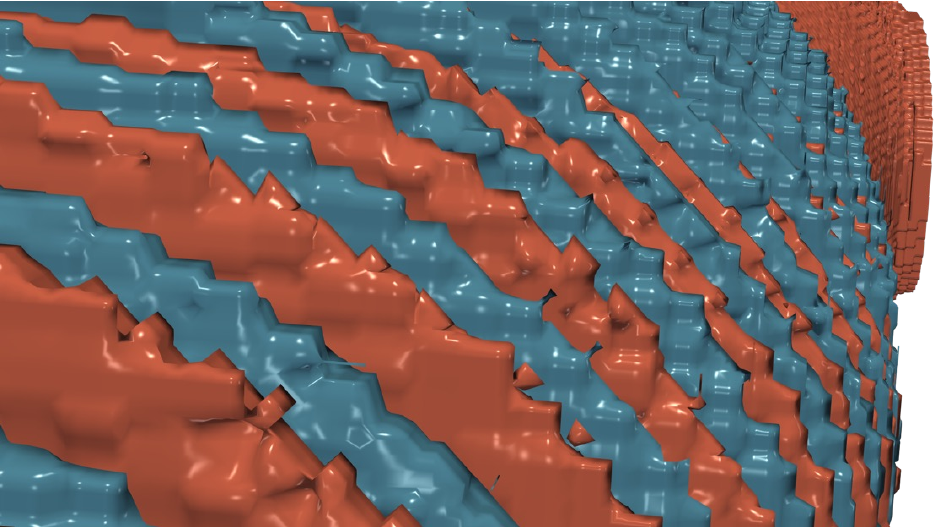}
    \caption{ZFP, SSIM=.72, PSNR=75.5}
    \label{fig:post-iso-amric-2}
  \end{subfigure}
  \begin{subfigure}[t]{0.32\linewidth}
    \centering
    \includegraphics[width=\linewidth]{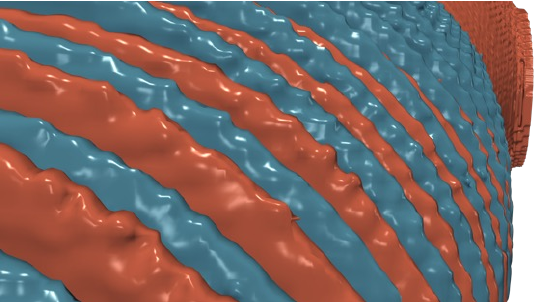}
    \caption{Processed ZFP, SSIM=.79, PSNR=78.1}
    \label{fig:post-iso-ours-2}
  \end{subfigure}
   \begin{subfigure}[t]{0.32\linewidth}
    \centering
    \includegraphics[width=\linewidth]{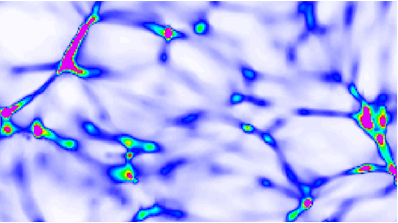}
    \caption[t]{Original data, Nyx}
    \label{fig:post-nyx-ori}
  \end{subfigure}\
  \begin{subfigure}[t]{0.32\linewidth}
    \centering
    \includegraphics[width=\linewidth]{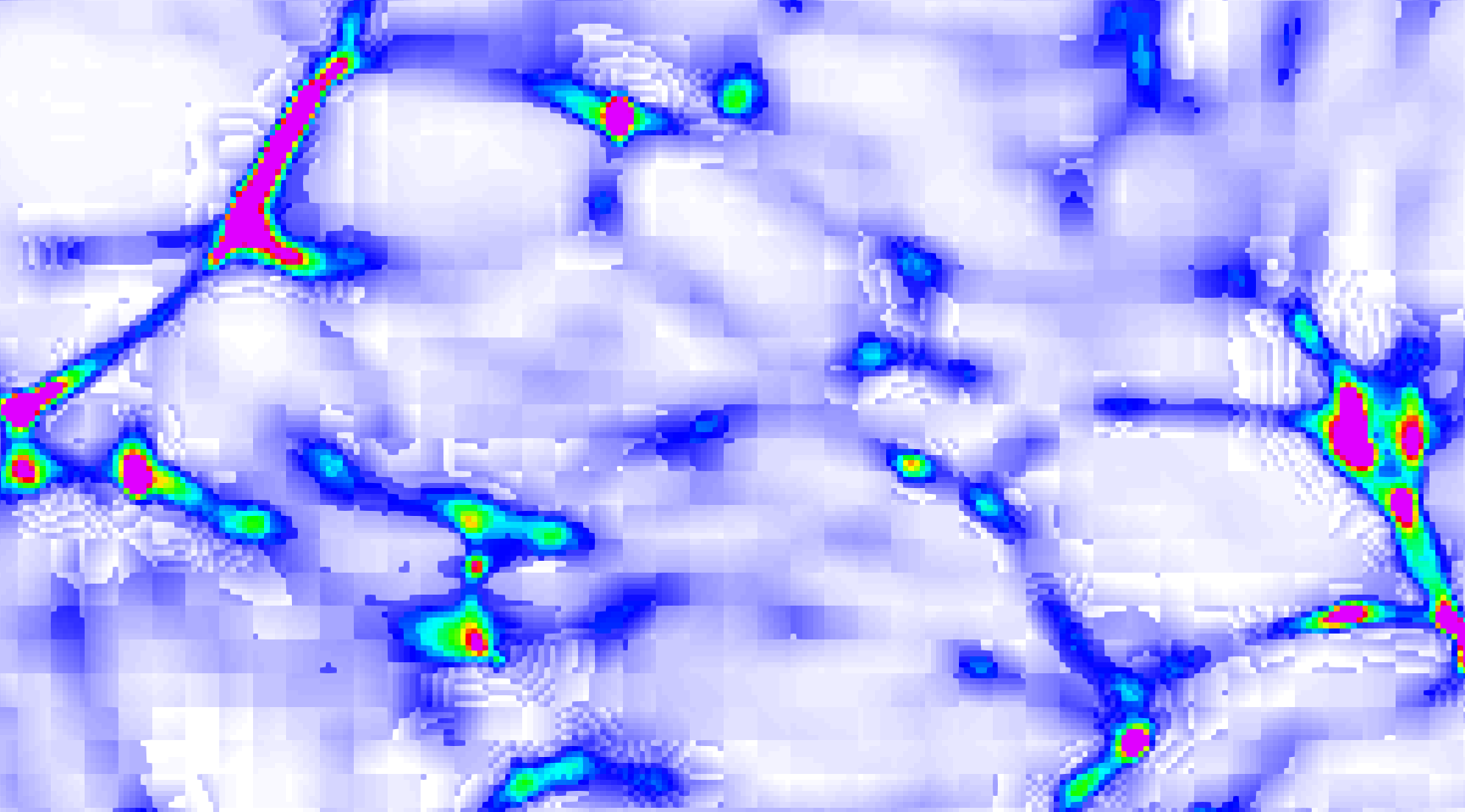}
    \caption{SZ2, SSIM=.76, PSNR=116.0}
    \label{fig:post-nyx-comp}
  \end{subfigure}
  \begin{subfigure}[t]{0.32\linewidth}
    \centering
    \includegraphics[width=\linewidth]{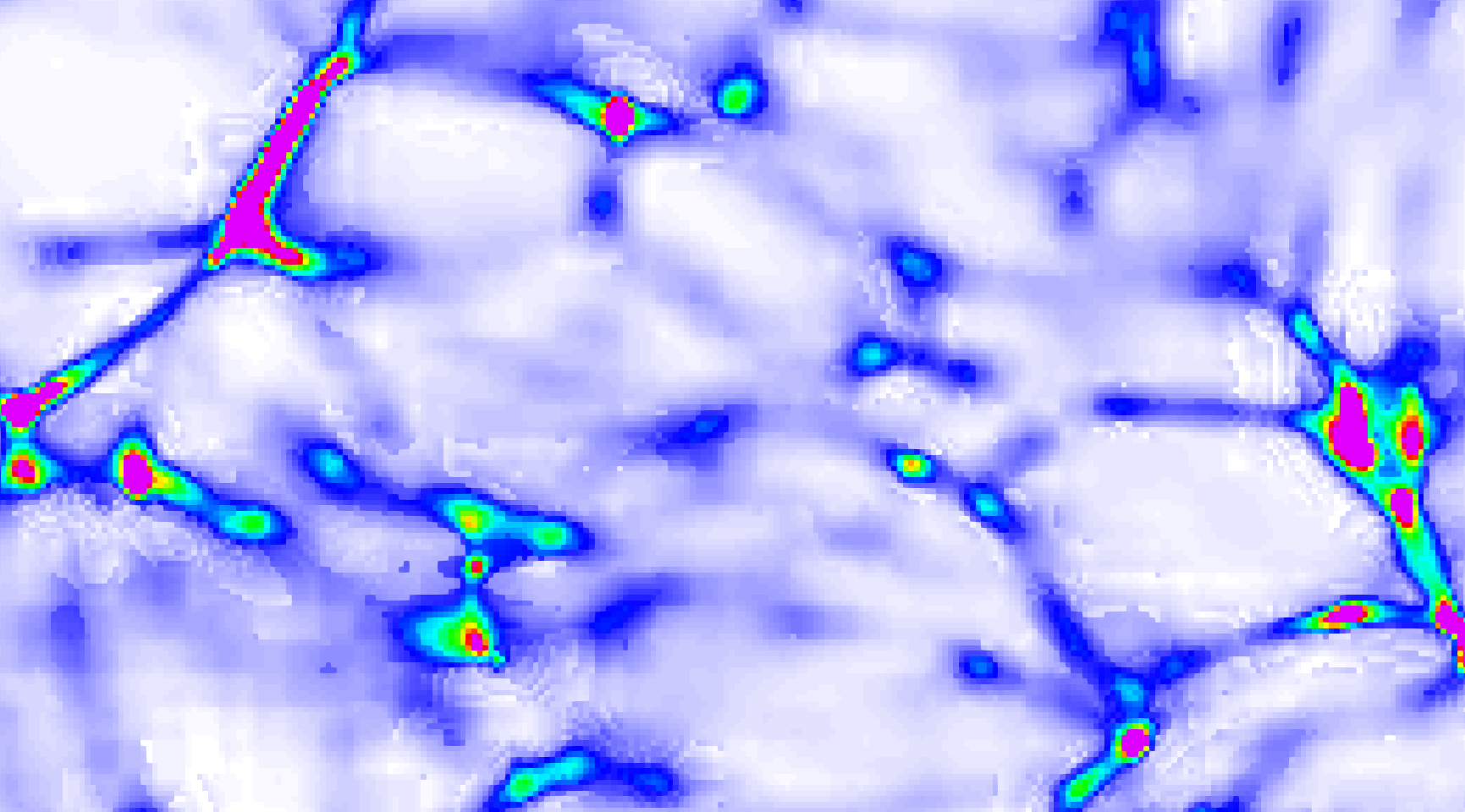}
    \caption{Processed SZ2, SSIM=.85, PSNR=118.1}
    \label{fig:post-nyx-post}
  \end{subfigure}

    \caption[t]{Visual comparison (iso-surface and 2D slice) of original data, decompressed data produced by ZFP and SZ2, and after our post-process on WarpX's ``Ez'' field and Nyx's ``density'' field. The CR is 139 and 143, respectively.}
    \label{fig:post-iso}
    \vspace{-1mm}
\end{figure*}

For block-wise scientific compressors like SZ2 and ZFP, previous studies have made significant strides in optimization for multi-resolution data. However, there remains scope for enhancement. Block-wise compressors often produce limited compression quality and are prone to compression artifacts~\cite{wang2023amrvis}, as shown in Fig.~\ref{fig:post-iso-amric-2}. To address these issues, we introduce a fast and effective post-processing solution that enhances the data quality of block-wise compressors. As shown in Fig.~\ref{fig:post-iso-ours-2}, our post-processing solution significantly reduces compression artifacts and errors. We will now discuss the challenges posed by block-wise compressors and detail our post-processing approach.

\textbf{\textit{Challenge: low quality of block-wise compression.}}
The low-quality issue of block-wise compressors is mainly attributed to their block-wise nature of dividing the dataset into small blocks (e.g., $4 \times 4 \times 4$) before the compression.
Specifically, the partition can cause each block to lose spatial information of its neighboring blocks, losing the opportunity for better compression quality.
Furthermore, the separate processing of blocks disrupts the coherence of features that span across the block boundaries, leading to a degradation in data visualization and quality.
It is important to note that, for SZ2, the issue with blocking artifacts will be more severe for multi-resolution data than for uniform-resolution data. This is because, for multi-resolution data, SZ2 needs to reduce its compression block size from $6 \times 6 \times 6$ to $4 \times 4 \times 4$ to achieve optimal performance~\cite{amric}, thus leading to more artifacts due to the smaller block size.

\begin{table}
  \centering\sffamily\scriptsize
  \caption{Comparison of data quality (in PSNR) of original decompressed data from ZFP,  decompressed data processed by image smooth/denoise filters, and our solution.}
  \label{tab:vsimg}
  \begin{tabular}{@{}cccccc@{}}
    \toprule
         & \begin{tabular}[c]{@{}c@{}}Decomp. \\ data\end{tabular} & \begin{tabular}[c]{@{}c@{}}Median \\ Filter\end{tabular} & \begin{tabular}[c]{@{}c@{}}Gaussian \\ Blur\end{tabular} & \begin{tabular}[c]{@{}c@{}}Anisotropic \\ Diffusion\end{tabular} & \textbf{Ours} \\
    \midrule
    PSNR & 80.5                                                    & 67.2                                                     & 71.6                                                     & 74.4                                                             & \textbf{82.9} \\
    \bottomrule
  \end{tabular}
\end{table}

\textbf{\textit{Limitation of the image processing filters.}}
Numerous image smoothing and denoising techniques, such as Anisotropic Diffusion, Gaussian Blur, and Median Filter, are widely used for post-processing. However, their effectiveness often diminishes when applied to decompressed data from error-bounded scientific compressors. This shortfall arises because these filters are designed for lossy image compressors like JPEG. When used on scientific data, they can over-smooth the data, leading to significant detail loss and a marked reduction in PSNR, as illustrated in TABLE~\ref{tab:vsimg}. This issue stems from the filters' lack of consideration for the error-bounded nature of the decompressed data, resulting in a notable deviation from the original dataset.

\begin{figure}
   \centering

  \begin{minipage}[c]{0.67\linewidth}
    \includegraphics[width=\linewidth]{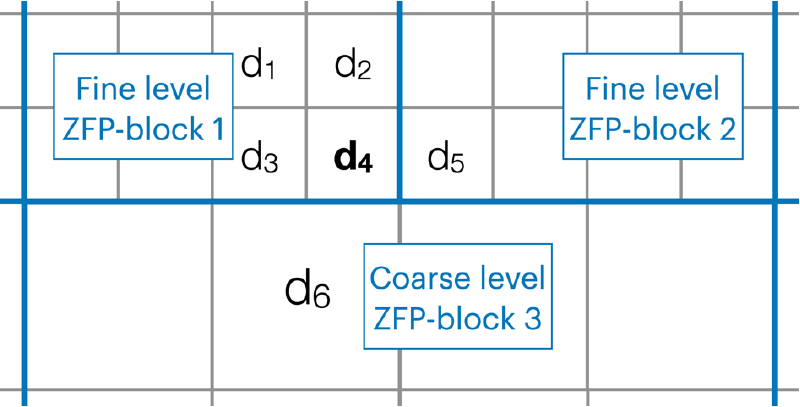}
  \end{minipage}\enspace~%
    \caption{Example of the multi-resolution data gird when using ZFP compressor, the gray grid indicates data points and the bold blue box indicates $4\times4$ blocks partitioned by ZFP.}
    \label{fig:grids}
\end{figure}

\textit{\textbf{Improvement: Use post-process to improve the compression quality.
  }}To tackle this challenge, we introduce an adaptive post-processing technique specifically designed for error-bounded scientific lossy compressors. This method starts by applying B\'ezier curves to exchange spatial information overlooked during compression among data blocks. It then utilizes the error-bound properties of the decompressed data and dynamically adjusts the processing intensity. This strategy markedly improves both visualization quality (e.g., Structural Similarity Index Measure (SSIM)) and data quality (e.g., PSNR) of the decompressed output.
Particularly, it excels at mitigating blocking artifacts, a common drawback of block-wise compression.

We opt for B\'ezier curves due to their ability to smooth transitions between points, effectively mitigating discontinuities or artifacts introduced during compression. Additionally, B\'ezier curves are computationally efficient and highly parallelizable, making them suitable for post-processing needs where computational speed is important.

\begin{figure}
  \centering
  \begin{minipage}{.68\linewidth}
    \includegraphics[width=\linewidth]{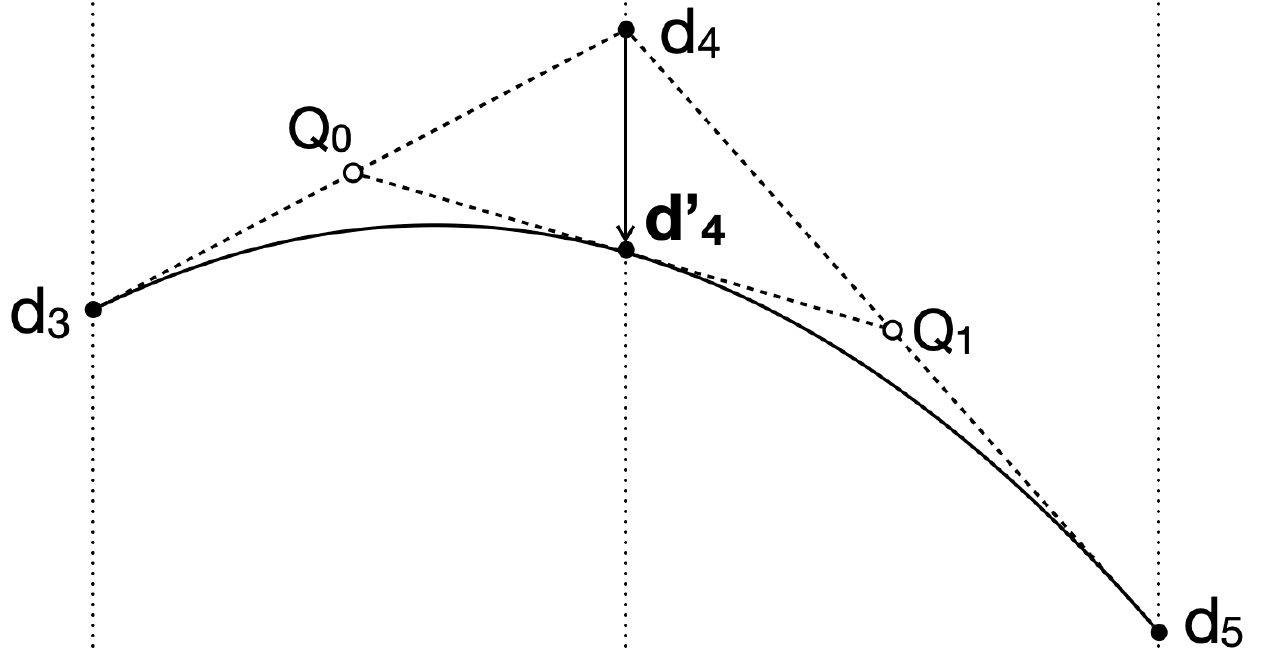}
  \end{minipage}~%
    \caption{Example of the B\'ezier curves for $t=0.5$, $Q_0$ and $Q_1$ are the midpoints of $d_3$$d_4$ and $d_4$$d_5$. $d_4'$ is obtained by B(0.5), mid of $Q_0$$Q_1$.}
    \label{fig:bzr}
  \vspace{-2mm}
\end{figure}

A 2D example is illustrated in Fig.~\ref{fig:grids} using ZFP. %
Data points are partitioned into $4\times4$ blocks for compression, isolating them from points in other blocks.
We aim to utilize B\'ezier curves to exchange spatial information between adjacent data blocks, thereby enhancing data quality.
Specifically, for decompressed data at the boundary (e.g., $d_4$), we can leverage its neighboring point $d_5$ along the x-direction from another block to improve its quality. This is achieved by constructing a quadratic B\'ezier curve with $d_3$, $d_4$, and $d_5$, where $d_3$ and $d_5$ are the start and end points, respectively, and $d_4$ serves as the control point. The curve is defined as:
\[
  B(t) = (1-t)^2d_3 + 2(1-t)td_4 + t^2d_5 \quad \text{for} \quad 0 \leq t \leq 1
\]
with $t$ being the parameter ranging from 0 to 1. As $t$ progresses from 0 to 1, the B\'ezier curve formula generates points tracing the curve's path from $d_3$ to $d_5$. The adjusted $d_4'$ is derived at $t=0.5$ ($d_4'~=~B(0.5)$) as shown in Fig.~\ref{fig:bzr}.
This B\'ezier curve approach is applied for each dimension separately and can be easily extended to multi-resolution scenarios. For instance, in the y direction, $d_2$ and $d_6$ would be used to process $d_4$.

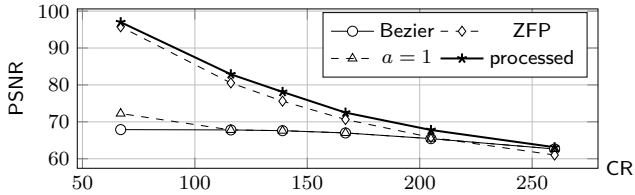
\begin{figure}

  \sffamily
  \centering\footnotesize
  \resizebox{0.98\linewidth}{!}{
  \begin{tikzpicture}
    \begin{axis}[width=\linewidth, height=1.5in,
        ylabel=PSNR, ylabel near ticks,
        xlabel=CR, xlabel near ticks, xlabel style={at=(current axis.right of origin), anchor=west},
        legend columns=2,
        cycle list name=MarkerCycleList,
        grid]
      \addplot coordinates {  %
          (260.0,62.645639)(205.0,65.448264)(167.0,66.993832)(139.0,67.605028)(116.0,67.810216)(67.0,67.896396)
        };
      \addplot coordinates {  %
          (260.0,61.016423)(205.0,65.788489)(167.0,70.636122)(139.0,75.593518)(116.0,80.538009)(67.0,95.755996)
        };
      \addplot coordinates {  %
          (260.0,62.645639)(205.0,65.448264)(167.0,66.993832)(139.0,67.605028)(116.0,67.810218)(67.0,72.249463)
        };
      \addplot[mark=star, thick] coordinates {  %
          (260.0,63.12682)(205.0,67.792894)(167.0,72.499084)(139.0,78.101964)(116.0,82.88392)(67.0,96.992286)
        };
      \legend{Bezier,ZFP,$a=1$,processed}

    \end{axis}
  \end{tikzpicture}
    }
  \caption{ Rate-distortion comparison of different post-process approaches on WarpX using ZFP.}
  \label{fig:post-wpx-zfp-off}

\end{figure}

\textit{\textbf{Leverage the error-bounded feature in decompressed data.}}
However, neglecting the error-bounded nature of decompressed data can significantly reduce the quality of the process, as illustrated in Fig.~\ref{fig:post-wpx-zfp-off}. Sole dependence on the B\'ezier curve (represented by ``\textbf{\textit{Bezier}}'') severely impacts the quality of ZFP decompressed data, mirroring the limitations encountered with image filters.
For an error-bounded compressor, the decompressed data point $d_4$ must stay within the error bounds $eb$ of the original data $o_4$. Therefore we have: $o_4 \in [d_4 ~-~ eb, d_4 ~+~ eb]$.
This condition suggests that when processing $d_4$ to $d_4'$, $d_4'$ should fall within $[d_4 - eb, d_4 + eb]$, guaranteeing:
\[
  d_4' = \max(\min(B(0.5), d_4 + eb), d_4 - eb)
\]
This formula ensures that $d_4'$ remains within the error limits, maintaining the decompressed data's integrity.

\begin{figure}[h]
  \centering
  \includegraphics[width=0.92\columnwidth]{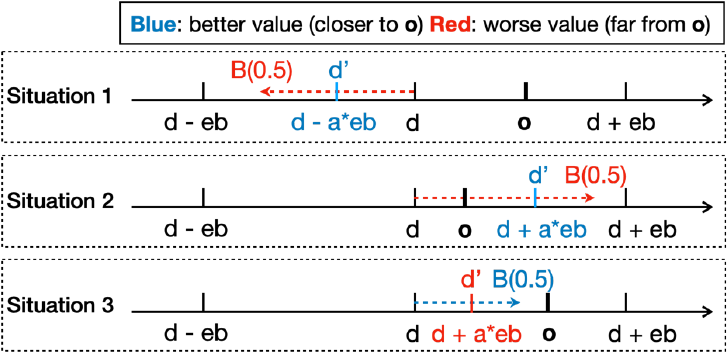}
  \caption{Example of the impact of setting smaller limit/intensity of the post-process under different situations, blue color indicates better post-process outcome and red color indicates worse.}
  \label{limit}
\end{figure}

\textbf{\textit{Further improve the process quality using dynamic limit/intensity.}}
Nevertheless, utilizing the error-bound information is still insufficient for achieving optimal post-processing quality. To enhance the data quality, we must adaptively limit the actual error bound used in the post-process, $eb'$ ($eb' = a \cdot eb$, $a<1$), making it smaller to control the post-process intensity.

To clarify the necessity of a smaller $eb'$, we'll examine a sample scenario. Assume the original data $o$ is larger than the decompressed data $d$. Initially, as illustrated at the top of Fig.~\ref{limit} (situation 1), there might be cases where the B\'ezier curve predicts in the opposite direction ($B(0.5) < d$). In such instances, a small $a$ helps prevent $d'$ from deviating excessively from original data $o$.
Secondly, as depicted in the middle of Fig.~\ref{limit} (situation 2), when the B\'ezier curve accurately predicts but overshoots beyond the original data ($B(0.5) > o$), a smaller $a$ helps ensure $d'$ remains closer to $o$.
However, if the B\'ezier curve correctly predicts within the bounds ($o < B(0.5) < d$), overly reducing $a$ prevents $d'$ from getting closer to $o$, as shown in the bottom of Fig.~\ref{limit} (situation 3).

We seek to maximize the gain from post-processing:
\[
  h \cdot (|e| - |e'|) - (1 - h) \cdot (|e'| - |e|)  \quad 0 \leq h \leq 1
\]
where \(h\) denotes the rate at the B\'ezier curve does not make opposite predictions. \(e\) and \(e'\) denote the compression errors before and after post-processing, which can be reformulated as:
\[
  \underset{a}{\text{maximize}} \quad h \cdot (|o - d| - |o - d'|) - (1 - h) \cdot |d' - d|,
\]
by adjusting \(a\), under the constraint:
\[
  d' = \max(\min(B(0.5), d + a \cdot eb), d - a \cdot eb) \quad 0 \leq a \leq 1.
\]

However, finding the optimal $a$ analytically is not feasible due to the presence of absolute values in the objective function and the piece-wise definition of $d'$.
Moreover, obtaining necessary parameters before compression—like the hit rate $h$ also incurs additional costs. Thus, we employ a sampling-based numerical optimization approach to iteratively find the optimal \(a\).

Now we present how the optimal \(a\) is dynamically determined through a lightweight compression sampling process.
Extensive experimentation across various datasets has enabled us to refine our selection of the best candidate parameters for our algorithm. Specifically, for SZ2, \(a_{\text{sz}}\) is narrowed to the set \(\{0.05, 0.1, 0.015, ..., 0.45, 0.5\}\), and for ZFP, \(a_{\text{zfp}}\) is set to \(\{0.005, 0.01, 0.015, ..., 0.05\}\).
These values can achieve optimal or near-optimal performance in most cases, while also being practical for evaluation.
The candidate for ZFP is smaller due to its underestimation characteristic, which leads to a smaller max real compression error than the given error bound.

\begin{table}
  \centering
  \caption{Rate-distortion comparison of original decompressed data and our post-process approach on WarpX using SZ2.}
  \label{tab:post-wpx-sz-off}

  \resizebox{\linewidth}{!}{%
    \begin{tabular}{@{}l *{7}{r} @{}}
      \toprule
      CR           & \textit{273 }  & \textit{ 207} & \textit{153}   & \textit{ 126 }  & \textit{104}  & \textit{ 62}    & \textit{34}    \\
      PSNR-SZ2     & 67.8           & 72.8          & 79.6           & 84.8            & 90.0          & 101.9           & 114.4          \\
      PSNR-Proc'ed & \textbf{ 69.8} & \textbf{74.6} & \textbf{81.1 } & \textbf{ 86.2 } & \textbf{91.2} & \textbf{102.6 } & \textbf{114.9} \\ %
      \bottomrule
    \end{tabular}%
  }
\end{table}

Our methodology starts by sampling \(i^3\) data blocks of size \((j \times \text{blocksize})^3\), where \(\text{blocksize}\) refers to the compressor's block-wise compression size. We aim for a sampling rate below 1.5\%, sufficient for identifying the optimal \(a\) with minimal overhead.
Then, for each dimension, we utilize stochastic gradient descent (SGD) to find the optimal \(a\) from the candidates that minimize the overall norm2 compression error.

Fig.~\ref{fig:post-wpx-zfp-off} shows that our post-processing with dynamic limit/intensity (denoted by \textbf{\textit{``Process''}}) significantly enhances ZFP decompressed data quality. The \textbf{\textit{``a=1''}} curve represents performance without the dynamic limit, showing low performance. Fig.~\ref{fig:post-iso} clearly illustrates how our post-processing significantly enhances data quality through visualization. In addition to ZFP, TABLE~\ref{tab:post-wpx-sz-off} illustrates our approach's effectiveness in improving SZ2's compressed data quality.
Furthermore, our post-processing can also improve the data quality for global compressors like SZ3 in multi-resolution scenarios. Because multi-resolution data need to be partitioned before compression, as discussed in \S\ref{sec:pad}. Detailed performance outcomes will be shown in \S\ref{sec:eva-ist}.

\subsection{Uncertainty Visualisation for Compression}
\label{sec:unc}

In this work, we employ uncertainty visualization to examine the effects of compression on data.
Specifically, we explore how compression errors influence the positions of isosurfaces, which are highly sensitive to errors and can be significantly altered by compression-related inaccuracies. This sensitivity provides a valuable perspective for deepening our understanding of compression's impact on data.

Multiple previous contributions have studied the impact of uncertainty in data on isosurface visualization~\cite{TA:Pothkow:2011:probMarchingCubes, athawale2015isosurface,TA:Athawale:2021:topoMappingUncertaintyMarchingCubes, TA:2016:Ferstl:isosurfaceUncertainty}. In this work, we leverage the probabilistic marching cubes idea~\cite{TA:Pothkow:2011:probMarchingCubes,TA:Athawale:2021:topoMappingUncertaintyMarchingCubes,wang2023funmc} to gain insight into the effect of compression errors on isosurface positions. The probabilistic marching cubes algorithm models per-voxel error as a probability distribution to derive the spatial probability distribution of isosurfaces.
Our primary objective is to utilize the error distribution of decompressed data to analyze isosurface uncertainty.
In both ZFP and SZ compressed data, errors follow a normal distribution~\cite{zfp-dist}, especially when the error bound is large~\cite{sian22icde}.

Thus, we focus on the normal distribution in this work, given our focus on cases with larger error bounds. Modeling uncertainty per voxel as a normal distribution involves determining the mean and variance of the uncertainty (compression error) per voxel, which is challenging because the error information is lost after compression. However, as illustrated in Fig.~\ref{fig:overview}, we sample the compression error during the compression process for post-processing needs. This sampled compression error can also be used to obtain the mean and variance of the error with minimal overhead by reusing the information.

\textbf{\textit{Isovalue related variance.}}
Given the fact that the data points close to the isovalue are more likely to be considered for the isosurface construction.
When computing the variance, we focus on data points with values near the isovalue instead of using all the sampled points. This approach allows for a more accurate variance calculation for the given isovalue, as the compression error could depend on the data value.

\begin{figure}
  \centering
  \begin{subfigure}[t]{0.32\linewidth}
    \centering
    \includegraphics[width=\linewidth]{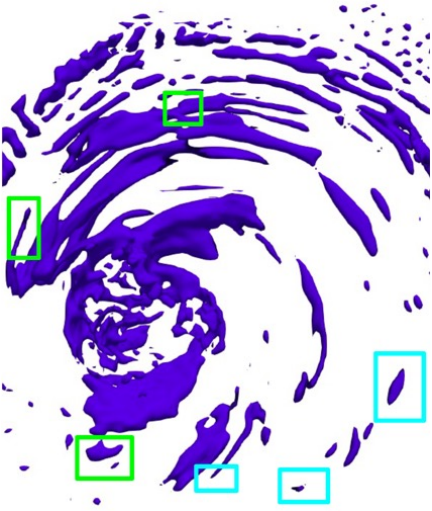}
    \caption[t]{Original}
    \label{fig:unc-ori}
  \end{subfigure}
  \begin{subfigure}[t]{0.32\linewidth}
    \centering
    \includegraphics[width=\linewidth]{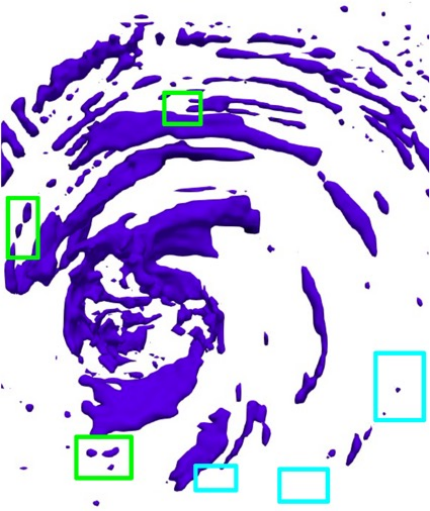}
    \caption{Decompressed}
    \label{fig:unc-comp}
  \end{subfigure}
  \begin{subfigure}[t]{0.32\linewidth}
    \centering
    \includegraphics[width=\linewidth]{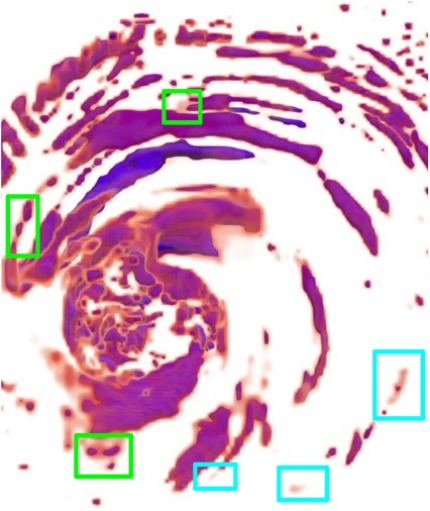}
    \caption{W/ Uncertainty}
    \label{fig:unc-both}
  \end{subfigure}
  \caption[t]{Vis of original data, decompressed data (generated by our workflow using ZFP, CR = 240), and decompressed data with uncertainty, cyan/green box highlights the missing/cracking isosurface.}
  \label{fig:unc}
\end{figure}

Having characterized uncertainty with error distribution near isovalue in decompressed data, we apply the probabilistic marching cubes techniques~\cite{TA:Pothkow:2011:probMarchingCubes,TA:Athawale:2021:topoMappingUncertaintyMarchingCubes} to gain insight into spatial uncertainty in isosurface arising from compression.
\textcolor{black}{Fig.~\ref{fig:unc} illustrates how uncertainty visualization helps in understanding the error in decompressed data.}
\textcolor{black}{Specifically, Fig.~\ref{fig:unc-ori} and Fig.~\ref{fig:unc-comp} visualize the isosurfaces for the Hurricane dataset~\cite{hurricane-data} extracted from the original data and decompressed data, respectively. Fig.~\ref{fig:unc-both} visualizes uncertainty in red using our approach for the decompressed data. The boxes in Fig.~\ref{fig:unc} highlight the topological features that are missed/broken in visualization without uncertainty (Fig.~\ref{fig:unc-comp}) but are successfully recovered by the one with uncertainty visualization (Fig.~\ref{fig:unc-both}). 
For example, the cyan boxes illustrate features that disappear from the original data in Fig.~\ref{fig:unc-ori} because of the compression errors, but whose potential presence is denoted by the red regions in Fig.~\ref{fig:unc-both}. Thus, the visualization of spatial uncertainty mitigates data misrepresentation arising from compression errors. }

This phenomenon occurs because the isosurface is prone to being pruned due to compression errors, attributed to its binary nature.
A moderate compression error can cause the isosurface to disappear completely; for example, if all the corresponding data values fall below the isovalue after compression. On the other hand, the isosurface uncertainty visualization, as described in~\cite{TA:Pothkow:2011:probMarchingCubes,TA:Athawale:2021:topoMappingUncertaintyMarchingCubes}, employs a more informative approach. It enhances visualization by incorporating the uncertainty (i.e., error distribution) of the decompressed data, rather than solely considering the decompressed data itself.

\section{Experimental Evaluation}
\subsection{Experimental Setup.}
\textbf{Applications and datasets.}
We conducted both in-situ and offline experiments. For the in-situ experiments, we selected two real-world applications: the Nyx cosmology simulation~\cite{nyx} and the WarpX electromagnetic
simulation~\cite{warpx,warpx-gordon}.
These were conducted on the Bridges-2~\cite{Bridges-2, bridges-2-paper}, where each node is equipped with two AMD EPYC 7742 CPUs and 256 GB RAM. Our experiments utilized 128 cores. Nyx serves as an AMR application, fully supporting AMR features. WarpX is utilized for experiments involving adaptive data (derived from uniform-resolution data) as WarpX does not yet fully support AMR.

In addition to in-situ evaluation, we also evaluated our solution using five different offline datasets from four distinct applications to demonstrate our solution's broad applicability. The offline evaluation included multi-resolution data with different resolution levels and density (density refers to the proportion of data within the entire domain) and uniform-resolution data as specified in TABLE~\ref{tab:dataset}. Specifically, we tested the Rayleigh-Taylor (denoted as``RT'') dataset generated by the IAMR fluid dynamics simulation~\cite{IAMR}, the S3D combustion simulation, the Hurricane Isabel dataset~\cite{sdrbench}, and two additional Nyx datasets (denoted as ``T2'' and ``T3'') from different timesteps.

\begin{table}[h]
    \setlength{\columnsep}{0pt}
    \caption{Our tested datasets}
    \label{tab:dataset}
    \centering\sffamily\scriptsize
    \newcommand{\MRESO}[2]{\makebox[4.5em][l]{#1}\makebox[7em][l]{#2}}

\begin{tabular}{@{} l llr @{}}
    \toprule
    \textbf{Dataset} & \textbf{Property}
                     &
    \textbf{\begin{tabular}[c]{@{}l@{}}(Size, Density) per Level\\ Fine to Coarse\end{tabular}}
                     &
    \textbf{\begin{tabular}[c]{@{}c@{}}Per-Timestep\\Data Size\end{tabular}}
    \\ \midrule
    {Nyx-T1}         & {In-situ, AMR}    & \MRESO{fine:}{$(512^3,18\%)$}              & 3.1 GB\quad\null \\
                     & {2 levels}        & \MRESO{coarse:}{$(256^3,82\%)$}                               \\
    \cmidrule(l){2-4}
    {WarpX}          & {In-situ, Adpt}   & \MRESO{fine:}{$(256^2\times2048, 50\%)$}   & 6.3 GB\quad\null \\
                     & {2 levels}        & \MRESO{coarse:}{$(128^2\times1024, 50\%)$}                    \\
    \cmidrule(l){2-4}
    {RT}             & {Offline, AMR}    & \MRESO{finest:}{$(512^3,15\%)$}            & 2 GB\quad\null   \\
                     & {3 levels}        & \MRESO{medium:}{$(256^3,31\%)$}                               \\
                     &                   & \MRESO{coarse:}{$(128^3,54\%)$}                               \\
    \cmidrule(l){2-4}
    {Nyx-T2}         & {Offline, AMR}    & \MRESO{fine:}{$(512^3,58\%)$}              & 7.1 GB\quad\null \\
                     & {2 levels}        & \MRESO{coarse:}{$(256^3,42\%)$}                               \\
    \cmidrule(l){2-4}
    {Hurri}          & {Offline, Adpt}   & \MRESO{fine:}{$(500^2\times100, 35\%)$}    & 1.1 GB\quad\null \\
                     & {2 levels}        & \MRESO{coarse:}{$(250^2\times50, 65\%)$}                      \\
    \cmidrule(l){2-4}
    {Nyx-T3}         & {Offline, Uni}    & $(512^3,100\%)$                            & 10 GB\quad\null  \\
    \cmidrule(l){2-4}
    {S3D}            & {Offline, Uni}    & $(512^3,100\%)$                            & 11 GB\quad\null  \\
    \bottomrule
\end{tabular}
\end{table}

\textbf{Comparison baseline.}
We evaluate our SZ3MR on both AMR data and adaptive data generated from uniform-resolution data.
In terms of AMR data, we benchmark our improved approach against AMRIC's SZ3 (referred to as ``AMRIC-SZ3'' \cite{amric}) and TAC's SZ3 (referred to as ``TAC-SZ3'', only for offline evaluation as it lacks an in-situ option \cite{wang2022tac}), and the original SZ3 (denote as ``Baseline-SZ3''). For adaptive data, our improved SZ3 is evaluated against the original SZ3 as TAC and AMRIC do not offer SZ3 implementation for adaptive data.

Further, we first conduct offline evaluations on our adaptive post-processing technique utilizing both SZ2 and ZFP across multi-resolution and uniform datasets. It is important to note that we employ AMRIC's SZ2 for multi-resolution data due to its superior compression capabilities compared to zMesh \cite{zMesh} and TAC.
Additionally, we have integrated our post-processing technique into the AMR application Nyx for in-situ evaluation, demonstrating that our approach significantly enhances the data quality of AMRIC-SZ2 through post-processing.

\subsection{In-situ Evaluation}
\label{sec:eva-ist}
\begin{figure}[h]\vspace{-3mm}
    \sffamily\centering\footnotesize
    \begin{tabular}{@{}c@{}}
    \begin{tikzpicture}
        \begin{axis}[
                title={Fine level, density=18\%},
                title style={yshift=-5pt},
                ,legend entries=
                    {Baseline-SZ3, AMRIC-SZ3, Ours (pad), Ours (pad+eb), Ours (processed)},
                ,legend to name=named,
                name=fig15a,
                width=2.0in, height=1.5in,
                xmin=0,
                xmax=310,
                ylabel=PSNR, ylabel near ticks, ylabel shift =-5pt,
                xlabel=CR, xlabel near ticks, xlabel style={at=(current axis.left of origin), anchor=east},
                legend columns=3,
                legend style={font=\scriptsize\strut,
                    },
                cycle list name=MarkerCycleList,
                grid]
            \addplot coordinates {  %
                    (165.2246557,54.559645)(142.06,56.670266)(112.95,59.68)(72.76,64.788204)(27.64532923,77.126416)(19.99033806,82.837084)
                };
            \addplot coordinates {  %
                    (200.96,54.83)(119.54,59.72)(119.54,59.72)(70.8,64.74)(26.1,77.03)(19.05,82.75)
                };
            \addplot [densely dotted,mark=star,mark options={solid}] coordinates {  %
                    (290.23,58.3)(186.83,60.89)(93.93,65.38)(28.36,76.99)(19.65,82.61)
                };
            \addplot [mark=triangle] coordinates {  %
                    (302.7513792,60.616817)(227.95,62.133469)(156.34,64.325559)(84.91,68.102059)(26.04,77.940988)(18.06,83.200611)
                };
            \addplot [densely dashed,thick,mark=asterisk,mark options={solid}]coordinates {  %
                    (302.7513792,61.456718)(227.95,62.935395)(156.34,64.997387)(84.91,68.535133)(26.04,78.072339)(18.06,83.263165)
                };
        \end{axis}

        \begin{axis}[
                title={Coarse level, density=82\%},
                title style={yshift=-5pt},
                name=fig15b,
                at=(fig15a.right of south east)
                anchor=left of south west,
                width=2.0in, height=1.5in,
                xshift=1.5em,
                xmin=0,
                xmax=210,
                legend columns=1,
                legend style={font=\tiny\strut,
                    },
                cycle list name=MarkerCycleList,
                grid]
            \addplot coordinates {  %
                    (98.00321354,28.266909)(78.75,30.946798)(63.49987573,33.521183)(30.15,43.077482)(24.33,46.48)(17.96,52.437574)
                };
            \addplot coordinates {  %
                    (142.32,27.99)(105.93,30.6)(80.22,33.13)(40.41,40.14)(26.8,45.8)(19.09,51.71)
                };
            \addplot [densely dotted,mark=star,mark options={solid}] coordinates {  %
                    (178.63,29.86)(116.12,31.97)(49.48,38.42)(29.78,43.73)(19.62,49.49)
                };
            \addplot [mark=triangle]coordinates {  %
            (197.3319002,32.000265)(141.8,33.989453)(104.2145739,35.75)(44.64,40.11)(26.62,44.48069)(17.59,49.954115)(15.76,51.838932)
            };
            \addplot [densely dashed,thick,mark=asterisk,mark options={solid}]coordinates {  %
                    (197.3319002,32.662127)(141.8,34.579789)(104.2145739,36.203136)(44.64,40.494232)(26.62,44.807773)(17.59,50.079349)
                };
        \end{axis}

    \end{tikzpicture}
    \\[-.8ex]
    \hspace{2.65em}\ref{named}%
\end{tabular}

    \caption{Rate-distortion comparison on AMR data of our SZ3MR approaches and baselines using Nyx AMR simulation (Nyx-T1).}
    \label{fig:nyx-ist}\vspace{-3mm}
\end{figure}
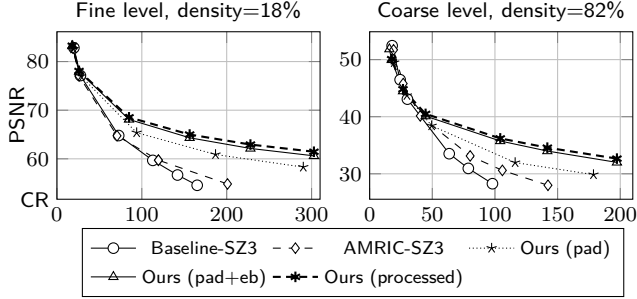

\textbf{\textit{In-situ Evaluation on AMR data compression.}}
As illustrated in Fig.~\ref{fig:nyx-ist}, our SZ3MR (with ``pad'' and ``eb'' detailing the performance of our two-step optimization) outperforms both the baseline and AMRIC across both refinement levels on Nyx, particularly at higher compression ratios. However, at the coarse level and with smaller compression ratios, our SZ3MR's performance is slightly worse than the baselines.
This is due to the high padding overhead given the smaller unit block size at the coarse level,
as discussed in \S\ref{sec:pad}.

We also compare the overall output time of our SZ3MR with that of AMRIC on Nyx.
The overall output time consists of (1) pre-processing (i.e., collecting data to the compression buffer) and (2) compression and writing the compressed data to the file system. As shown in TABLE~\ref{tab:nyx-time},
although our compression speed is slightly lower than AMRIC because of the padding overhead, our SZ3MR achieves a faster total output speed in both large and small error-bound settings.
The improvement is primarily attributed to our more efficient pre-processing stage, as AMRIC's stacking process is more complex and computationally intensive, requiring significant data rearrangement.

\begin{table}[h]
    \centering\sffamily\scriptsize
    \caption{Output time of AMRIC and our SZ3MR on Nyx-T1.}
    \label{tab:nyx-time}
    \begin{tabular}{@{}lcccc@{}}
        \toprule
        EB\_abs                         & \begin{tabular}[c]{@{}c@{}}Time\\ (Sec)\end{tabular} & Pre-process   & \begin{tabular}[c]{@{}c@{}}Comp. \& \\ Writing\end{tabular} & \begin{tabular}[c]{@{}c@{}}Total\\Time\end{tabular} \\
        \midrule
        \multirow{2}{*}{5.4E+9 (big)}   & AMRIC                                                & 1.22          & \textbf{1.62}                                               & 2.85                                                \\
                                        & Ours                                                 & \textbf{0.49} & 1.69                                                        & \textbf{2.18}                                       \\
        \midrule
        \multirow{2}{*}{2.7E+8 (small)} & AMRIC                                                & 1.23          & \textbf{2.30}                                               & 3.52                                                \\
                                        & Ours                                                 & \textbf{0.47} & 2.38                                                        & \textbf{2.85}                                       \\
        \bottomrule
    \end{tabular}
\end{table}

Our post-processing solution, as shown in TABLE~\ref{tab:nyx-post-ist}, significantly improves the quality of decompressed data for AMRIC-SZ2 on Nyx simulation at both resolution levels, with the degree of improvement being notably greater at higher compression ratios.
Furthermore, as outlined in \S\ref{sec:pad} and illustrated by the ``\textbf{Ours (processed)}'' curve in Fig.~\ref{fig:nyx-ist}, our post-processing also improves the data quality of SZ3 on multi-resolution data due to the need for partition. However, the improvement is less substantial than those achieved with block-wise compressors SZ2/ZFP.
This is because the partition size (unit block size) for multi-resolution data is larger than the block sizes used by SZ/ZFP (16 vs. 4), resulting in less room for improvement.

\begin{table}[h]
    \centering\sffamily\scriptsize
    \caption{Rate-distortion comparison of decompressed data and our post-process solution on both levels of Nyx-T1 using AMRIC-SZ2.}
    \label{tab:nyx-post-ist}
    \begin{tabular}{@{}lcccccc@{}}
        \toprule
        \multirow{3}{*}{Fine}   & CR             & \textit{270}  & \textit{165}  & \textit{113}  & \textit{73}   & \textit{28}   \\
                                & PSNR-AMRIC-SZ2 & 48.1          & 54.6          & 59.7          & 64.8          & 77.1          \\
                                & PSNR-Post-SZ2  & \textbf{50.1} & \textbf{56.9} & \textbf{61.8} & \textbf{66.5} & \textbf{77.6} \\
        \midrule
        \multirow{3}{*}{Coarse} & CR             & \textit{128}  & \textit{98}   & \textit{63}   & \textit{36}   & \textit{24}   \\
                                & PSNR-AMRIC-SZ2 & 25.3          & 28.3          & 33.5          & 40.7          & 46.5          \\
                                & PSNR-Post-SZ2  & \textbf{27.8} & \textbf{31.0} & \textbf{36.0} & \textbf{41.9} & \textbf{46.9} \\
        \bottomrule
    \end{tabular}
\end{table}

\begin{figure*}[t]

    \centering
    \begin{subfigure}[t]{0.32\linewidth}
        \centering
        \includegraphics[width=\linewidth]{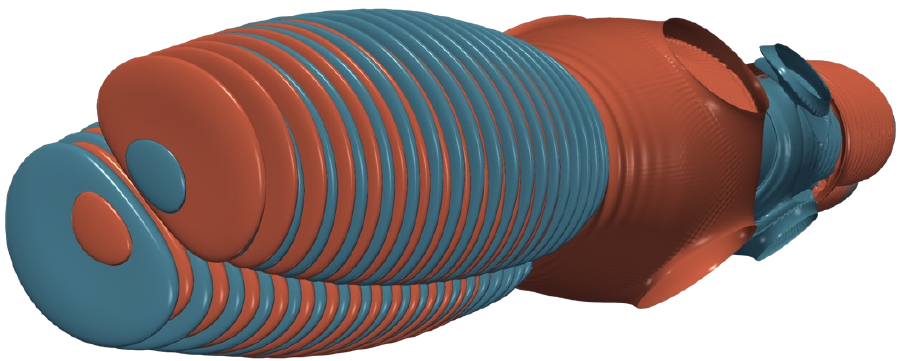}
        \caption[t]{Original data}
        \label{fig:wpx-sz3-ori}
    \end{subfigure}
    \begin{subfigure}[t]{0.32\linewidth}
        \centering
        \includegraphics[width=\linewidth]{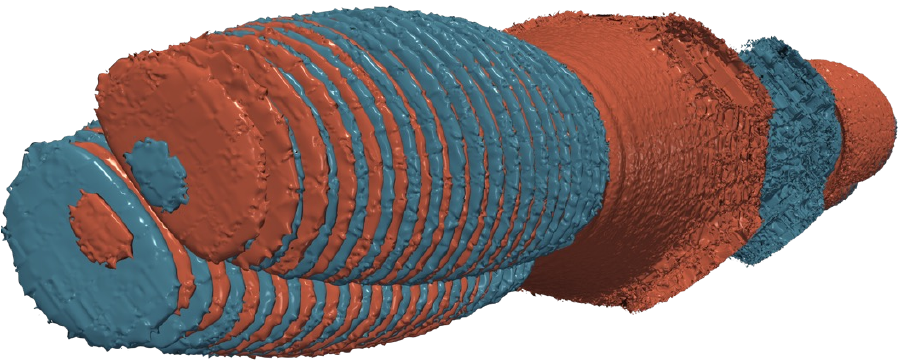}
        \caption{SZ3, CR=147, SSIM=0.662, PSNR=75.5}
        \label{fig:wpx-sz3-nast}
    \end{subfigure}
    \begin{subfigure}[t]{0.32\linewidth}
        \centering
        \includegraphics[width=\linewidth]{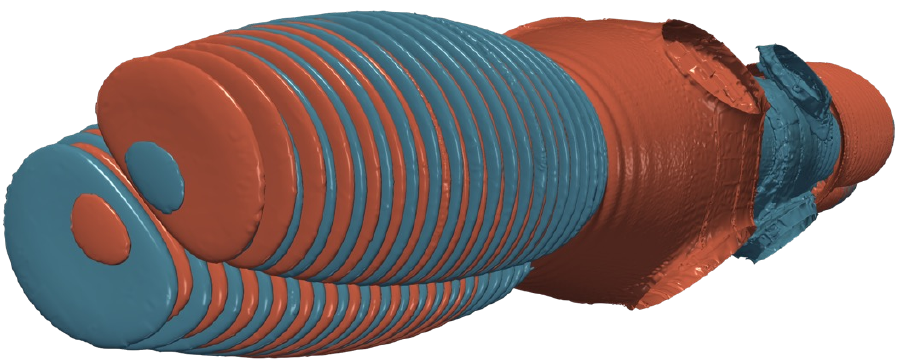}
        \caption{Ours, CR=147, SSIM=0.904, PSNR=86.9}
        \label{fig:wpx-sz3-ours}
    \end{subfigure}
    \caption[t]{Visual comparison (iso-surface) of original data and decompressed data produced by original SZ3 and our SZ3MR on WarpX (``Ez'' field).}
    \label{fig:warpx-iso}
    \vspace{-5mm}
\end{figure*}

\begin{figure}[h]
    \centering\sffamily\scriptsize
    \vspace{-2mm}
    \begin{tabular}{@{}c@{}}
    \begin{tikzpicture}
        \begin{axis}[
                title={WarpX},
                title style={yshift=-5pt},
                ,legend entries=
                    {Baseline-SZ3, Ours (pad), Ours (pad+eb)},
                ,legend to name=named,
                name=fig17a,
                width=2.0in, height=1.5in,
                ytick={70, 90, 110}, 
                xmin=0,
                xmax=310,
                ylabel=PSNR, ylabel near ticks, ylabel shift =-3pt,
                xlabel=CR, xlabel near ticks, xlabel style={at=(current axis.left of origin), anchor=east},
                legend columns=3,
                legend style={font=\scriptsize\strut,
                    },
                cycle list name=MarkerCycleList,
                grid]

            \addplot coordinates {  %
                    (245.9855077,64.075389)(193.4545006,69.235339)(147.1313417,75.503367)(110.8256579,82.532621)(90.53436656,87.80181)(51.17092561,101.344153)(40.29215389,106.648156)(31.42260878,112.226388)
                    (nan,nan)
                };
            \addplot [densely dotted,mark=star,mark options={solid}]coordinates {  %
                    (262.535882,77.159571)(163.8445094,83.836833)(116.245249,88.780272)(97.02537601,91.523465)(78.00684265,94.835413)(50.81737196,101.570987)(37.48052886,106.658149)
                    (27.74965944,112.06616)
                };
            \addplot[thick, mark=triangle] coordinates {  %
                    (288.4263836,78.43482)(226.744658,81.266574)(144.0570483,87.26233)(104.0508522,91.806206)(87.27918568,94.365227)(71.24586572,97.482885)(47.27460503,103.708465)(35.01217682,108.38509)
                    (26.16894308,113.440723)
                };

        \end{axis}

        \begin{axis}[
                title={Hurricane},
                title style={yshift=-5pt},
                name=fig17b,
                at=(fig17a.right of south east)
                anchor=left of south west,
                width=2.0in, height=1.5in,
                xshift=1.5em,
                ytick={50, 80, 110}, 
                xmin=0,
                xmax=230,
                legend columns=1,
                legend style={font=\tiny\strut,
                    },
                cycle list name=MarkerCycleList,
                grid]
            \addplot coordinates {  %
                    (206.754,42.5013)(158.098,46.4231)(123.809,51.3589)(102.763,54.506)(76.6215,59.5005)(60.7056,66.2339)(39.7534,80.232)(25.9139,99.3364)
                    (19.7014,113.412)
                };
            \addplot [densely dotted,mark=star,mark options={solid}]coordinates {  %
                    (217.379,46.9588)(160.457,51.9361)(129.999,55.0523)(94.002,59.9608)(71.6282,66.708)(45.0549,80.725)(28.8452,99.7194)(21.8109,113.836)
                    (nan,nan)
                };
            \addplot[thick, mark=triangle] coordinates {  %
                    (212.438,49.6866)(158.366,53.7442)(127.309,56.5281)(91.5038,62.5651)(70.2444,68.2707)(44.4333,82.1524)(28.4473,101.743)(21.4993,115.54)
                    (nan,nan)
                };
        \end{axis}

    \end{tikzpicture}
    \\[-.8ex]
    \hspace{2.65em}\ref{named}%
\end{tabular}
    \caption{Rate-distortion comparison on adaptive data of our SZ3MR and baselines using WarpX (in-situ) and Hurricane (offline) datasets.}
    \label{fig:warpx-hurricane-adp}
    \vspace{-3mm}
\end{figure}
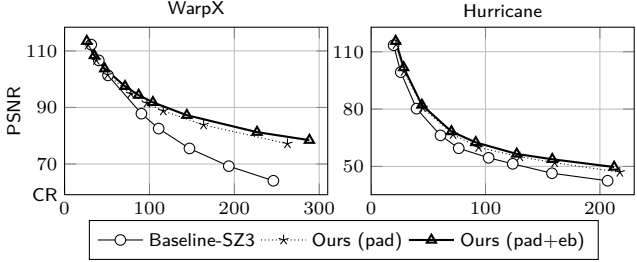

\textbf{\textit{In-situ evaluation on adaptive data compression.}}
Regarding adaptive data derived from uniform data, Fig.~\ref{fig:warpx-hurricane-adp} (left) shows our in-situ experiments with WarpX, demonstrating that our SZ3MR outperforms the original SZ3 baseline in most cases, except at lower compression ratios. It's important to note that AMRIC-SZ3 and TAC-SZ3 were not compared in this context due to their lack of support for adaptive data.
Furthermore, as shown in Fig.~\ref{fig:warpx-iso}, our SZ3MR notably enhances the compression quality (in terms of both PSNR and SSIM) and reduces visualization artifacts, offering a clear improvement over the baseline.

\subsection{Offline Evaluation}
\label{sec:eva-off}

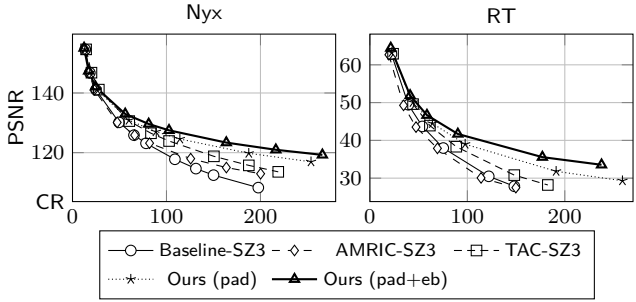
\begin{figure}[h]
    \footnotesize\centering\sffamily
    \begin{tabular}{@{}c@{}}
    \begin{tikzpicture}
        \begin{axis}[
                title={Nyx},
                title style={yshift=-5pt},
                ,legend entries=
                    {Baseline-SZ3, AMRIC-SZ3, TAC-SZ3, Ours (pad), Ours (pad+eb)},
                ,legend to name=named,
                name=fig18a,
                width=2.0in, height=1.5in,
                xmin=0,
                xmax=280,
                ylabel=PSNR, ylabel near ticks, ylabel shift =-3pt,
                xlabel=CR, xlabel near ticks, xlabel style={at=(current axis.left of origin), anchor=east},
                legend columns=3,
                legend style={font=\scriptsize\strut,
                    },
                cycle list name=MarkerCycleList,
                grid]

            \addplot coordinates {  %
                    (13.60061739,154.6131706)(18.43903134,146.7793925)(24.98679736,141.0479827)(49.50337416,130.1556088)(65.41833083,125.8660671)(78.23435185,123.1321498)(108.9578431,117.782322)(131.2956204,114.6594137)(150.0571668,112.4686816)(197.6990367,108.2841917)
                };
            \addplot coordinates {  %
                    (13.12637297,154.6130745)(17.67053944,146.7787679)(23.34304561,141.0436729)(47.56807383,130.1357603)(66.09939256,125.8741603)(82.00795874,123.186046)(125.771099,117.9964473)(163.7119095,115.0093487)(200.1645413,112.9350691)
                };
            \addplot [densely dashed,mark=square,mark options={solid}]coordinates {  %
                    (14.46979938,154.6157428)(20.16749419,146.8009641)(28.013418,141.1246102)(60.3626878,130.5819136)(83.50521583,126.5042286)(102.7623004,123.9022753)(150.4549415,118.7745465)(187.6158668,115.7406556)(218.9160939,113.5757484)
                };
            \addplot [densely dotted,mark=star,mark options={solid}]coordinates {  %
                    (13.0472678,154.6177227)(18.32601327,146.8170843)(25.90121189,141.1708627)(60.10400326,130.8744021)(88.79259144,127.0188968)(114.2513369,124.5786382)(187.5515381,119.7907876)(253.7878912,116.9356481)
                };
            \addplot[thick, mark=triangle] coordinates {  %
                    (12.33371609,155.0561023)(17.18086896,147.3910894)(24.46648676,142.0647266)(56.0835159,132.9728521)(81.06518814,129.6107424)(102.6995034,127.4855735)(163.650579,123.371779)(216.3488616,120.9685656)(265.8844711,119.2579275)
                };

        \end{axis}

        \begin{axis}[
                title={RT},
                title style={yshift=-5pt},
                name=fig18b,
                at=(fig18a.right of south east)
                anchor=left of south west,
                width=2.0in, height=1.5in,
                xshift=1.5em,
                xmin=0,
                xmax=270,
                legend columns=1,
                legend style={font=\tiny\strut,
                    },
                cycle list name=MarkerCycleList,
                grid]
            \addplot coordinates {  %
                    (22.13631438,62.84860314)(39.51187114,49.42300183)(53.89962931,43.6906848)(75.2855575,37.96949446)(121.9538452,30.48363189)(148.2171524,27.85037953)
                };
            \addplot coordinates {  %
                    (19.50021065,62.66628962)(34.53994637,49.27128922)(47.34102006,43.51547855)(69.30724228,37.88097898)(114.0802245,30.06163106)(149.7819232,27.49033263)
                };
            \addplot [densely dashed,mark=square,mark options={solid}]coordinates {  %
                    (23.77313224,62.930873)(44.27583741,49.60102415)(61.57073281,43.90732289)(88.63068773,38.32163227)(148.1469287,30.80444734)(182.5883263,28.19063264)
                };
            \addplot [densely dotted,mark=star,mark options={solid}]coordinates {  %
                    (22.44045532,63.24324306)(44.24841498,50.0460976)(64.1527251,44.4862188)(97.76757168,38.98536253)(191.1016036,31.80578047)(259.1524445,29.30898315)
                };
            \addplot[thick, mark=triangle] coordinates {  %
                    (21.42301043,64.44010034)(41.26562191,51.88883203)(58.55210616,46.6741149)(90.25321886,41.6929808)(176.9801744,35.56481692)(237.6878019,33.5356645)
                };
        \end{axis}

    \end{tikzpicture}
    \\[-.8ex]
    \hspace{2.65em}\ref{named}%
\end{tabular}
    \caption{Rate-distortion comparison on offline AMR data of our SZ3MR and baselines using Nyx-T2 and RT datasets.}
    \label{fig:Nyx-T2-RT-pad}
\end{figure}

\textbf{\textit{SZ3-MR on multi-resolution data.}}
As illustrated in Fig.~\ref{fig:Nyx-T2-RT-pad}, our method, after the two-step optimizations, outperforms all three baselines for both the Nyx-T2 and RT AMR datasets. It's observed that the AMRIC solution underperforms compared to the baseline on the RT dataset. We attribute this to the RT dataset having an additional refinement level compared to Nyx-T2, resulting in sparser data and more unsmooth boundaries due to merging non-adjacent blocks, leading to increased mispredictions.
Also, note that when the compression ratio is low, TAC yields slightly better performance than our solution on Nyx-T2, but its advantage is almost negligible on the RT dataset. This is because the RT has a smaller data size for each resolution level. Since TAC must compress the processed blocks with different shapes separately for each level (as mentioned in \S\ref{sec:pad}), the smaller data size will severe the encoding overhead issue of TAC and lead to a low compression ratio.

Regarding adaptive data derived from uniform data, as shown in Fig.~\ref{fig:warpx-hurricane-adp} (right), our adaptive error-bound solution offers limited enhancements until the high compression ratio. However, our padding technique consistently delivers significant improvements over the baseline across all compression ratios in the Hurricane dataset. We attribute this performance to the dataset's relative sparsity (i.e., numerous zero points), which enhances compressibility and offsets the padding overhead.

We also evaluated SZ3MR using application-specific power spectrum analysis on the Nyx-T2 dataset (see~\cite{jin2021adaptive,wang2022tac} for more details on power spectrum analysis in Nyx).
We compared the power spectrum $p'(k)$ of decompressed data with the original $p(k)$.
Typically, a maximum relative error threshold of 1\% is considered acceptable for all $k < 10$. 
Table~\ref{tab:ps} shows that under the same compression ratio, SZ3MR achieves a lower power spectrum error (including both the max and average errors for all $k < 10$) compared to all three baselines. Specifically, SZ3MR reduces the max power spectrum error by 75\%, 76\%, and 73\%, and reduces the average error by 74\%, 60\%, and 62\% compared to the original SZ3, AMRIC, and TAC, respectively, at the same compression ratio.

\begin{table}[h]
    \vspace{-1mm}
    \centering\sffamily\scriptsize
    \caption{Max and average power spectrum error comparison of our SZ3MR and baselines on Nyx-T2 under same CR for all $k < 10$.}
    \label{tab:ps}
    \begin{tabular}{@{}rcccc@{}}
        \toprule
        & Baseline-SZ3 & AMRIC-SZ3 & TAC-SZ3 & \textbf{Ours(pad+eb)} \\ 
        \midrule
        Avg Rel Error & 8.8E-03 & 5.7E-03 & 6.0E-03 & \textbf{2.3E-03} \\ 
        Max Rel Error & 2.7E-02 & 2.8E-02 & 2.5E-02 & \textbf{6.7E-03} \\ 
        \bottomrule
    \end{tabular}
    \vspace{-2mm}
\end{table}

\textbf{\textit{Post process for multi-resolution data.}}
As illustrated in  TABLE~\ref{tab:post-mr}, our post-process approach enhances the data quality in terms of PSNR for both the Hurricane and RT datasets across all compression ratios, with both SZ2 (optimized by AMRIC for multi-resolution data) and ZFP. Note that PSNR improvement is relatively modest at low compression ratios (e.g., under 30) because a lower CR indicates higher decompressed data quality, leaving limited room for improvement. When the compression ratio is low, our dynamic post-process approach can apply a conservative degree of post-processing intensity to ensure the original data quality remains uncompromised.

\begin{table}[h]
    \centering\sffamily\scriptsize
    \caption{Rate-distortion comparison of original decompressed data and our post-process approach on multiresolution datasets Hurricane and RT using ZFP and AMRIC-SZ2.}
    \label{tab:post-mr}
    \begin{tabular}{@{}lcccccccc@{}}
        \toprule
        \multirow{6}{*}{RT} & \multirow{3}{*}{ZFP} & CR        & \textit{184}  & \textit{143}  & \textit{118}  & \textit{72}   & \textit{43}   & \textit{27}   \\
                             &                      & PSNR-Ori  & 34.2          & 41.2          & 45.3          & 54.1          & 63.6          & 74.2          \\
                             &                      & PSNR-Post & \textbf{36.7} & \textbf{43.9} & \textbf{47.7} & \textbf{55.4} & \textbf{64.2} & \textbf{74.5} \\ \cmidrule{2-9}
                             & \multirow{3}{*}{SZ2} & CR        & \textit{257}  & \textit{180}  & \textit{122}  & \textit{75}   & \textit{40}   & \textit{22}   \\
                             &                      & PSNR-Ori  & 35.2          & 40.5          & 45.8          & 53.3          & 64.8          & 78.2          \\
                             &                      & PSNR-Post & \textbf{37.2} & \textbf{42.5} & \textbf{47.6} & \textbf{54.6} & \textbf{65.6} & \textbf{78.6} \\
        \midrule
        \multirow{6}{*}{Hur}  & \multirow{3}{*}{ZFP} & CR        & \textit{240}  & \textit{147}  & \textit{94}   & \textit{64}   & \textit{27}   & \textit{18}   \\
                             &                      & PSNR-Ori  & 40.1          & 43.7          & 47.8          & 52.6          & 68.5          & 80            \\
                             &                      & PSNR-Post & \textbf{42.1} & \textbf{45.6} & \textbf{49.5} & \textbf{53.8} & \textbf{69.2} & \textbf{80.5} \\ \cmidrule{2-9}
                             & \multirow{3}{*}{SZ2} & CR        & \textit{170}  & \textit{121}  & \textit{108}  & \textit{73}   & \textit{38}   & \textit{23}   \\
                             &                      & PSNR-Ori  & 41.9          & 44.3          & 45.3          & 49.9          & 62.4          & 75.8          \\
                             &                      & PSNR-Post & \textbf{43.2} & \textbf{45.9} & \textbf{47}   & \textbf{51.5} & \textbf{63.3} & \textbf{76.4} \\
        \bottomrule
    \end{tabular}%
\end{table}

\textbf{\textit{Post process for uniform resolution-data.}} Our post-processing method, as previously mentioned, demonstrates broad applicability, making it suitable for processing both uniform-resolution data and multi-resolution data from block-wise compressors.
As shown in TABLE~\ref{tab:post-uni}, and in alignment with our previous observations, our post-processing consistently enhances the data quality of the original SZ2 and ZFP outputs for both the uniform resolution datasets Nyx-T3 and S3D.

\begin{table}[h]
    \centering\sffamily\scriptsize
    \caption{Rate-distortion comparison of original decompressed data and our post-process approach on uniform resolution dataset S3D and Nyx-T3 using ZFP and SZ2.}
    \label{tab:post-uni}
    \resizebox{\linewidth}{!}{%
        \begin{tabular}{@{}ccccccccc@{}}
            \toprule
            \multirow{6}{*}{S3D} & \multirow{3}{*}{ZFP} & CR        & \textit{138}   & \textit{106}   & \textit{87}    & \textit{70}    & \textit{55}    & \textit{32}    \\
                                 &                      & PSNR-Ori  & 48.4           & 62.7           & 73.4           & 83.7           & 94             & 115.6          \\
                                 &                      & PSNR-Post & \textbf{51}    & \textbf{65.9}  & \textbf{75.4}  & \textbf{84.8}  & \textbf{94.7}  & \textbf{115.9} \\ \cmidrule{2-9}
                                 & \multirow{3}{*}{SZ2} & CR        & \textit{229}   & \textit{180}   & \textit{135}   & \textit{81}    & \textit{59}    & \textit{40}    \\
                                 &                      & PSNR-Ori  & 64.9           & 67.7           & 72.8           & 90.8           & 104.2          & 115.8          \\
                                 &                      & PSNR-Post & \textbf{67.5}  & \textbf{70.2}  & \textbf{74.7}  & \textbf{91.4}  & \textbf{104.4} & \textbf{116.0} \\ \midrule
            \multirow{6}{*}{Nyx} & \multirow{3}{*}{ZFP} & CR        & \textit{149}   & \textit{116}   & \textit{73}    & \textit{56}    & \textit{41}    & \textit{22}    \\
                                 &                      & PSNR-Ori  & 107.3          & 112.1          & 120.5          & 124.8          & 129.2          & 138.3          \\
                                 &                      & PSNR-Post & \textbf{109.3} & \textbf{114.2} & \textbf{122.8} & \textbf{126.8} & \textbf{130.9} & \textbf{139.2} \\ \cmidrule{2-9}
                                 & \multirow{3}{*}{SZ2} & CR        & \textit{214}   & \textit{143}   & \textit{94}    & \textit{53}    & \textit{34}    & \textit{13}    \\
                                 &                      & PSNR-Ori  & 112.5          & 116            & 119.7          & 124.8          & 128.9          & 140.3          \\
                                 &                      & PSNR-Post & \textbf{114.5} & \textbf{118.1} & \textbf{121.7} & \textbf{126.7} & \textbf{130.6} & \textbf{141.3} \\
            \bottomrule
        \end{tabular}%
    }
\end{table}

\textbf{\textit{Post process overhead.}}
Our post-processing solution is efficient and highly parallelizable, as mentioned in \S\ref{sec:post}, thereby introducing minimal overhead to the compression workflow. We employ OpenMP to accelerate our post-processing approach and assess its overhead using both SZ2 and ZFP, which are also optimized with OpenMP. It is important to note that using OpenMP with SZ2 can lead to a lower compression ratio due to the embarrassingly parallel. Thus, we have also conducted evaluations using the serial SZ2.
As demonstrated in the last column of TABLE~\ref{tab:post-time}, our post-processing introduces an overhead of only about 1.3\% for serial SZ2 and 3.5\% for SZ2/ZFP with OpenMP acceleration, respectively, under various compression ratios, utilizing 64 cores.

\begin{table}[h]
    \centering\sffamily\scriptsize
    \caption{Execution time of original compression workflow (columns 1 and 2) and post-processing (columns 3 and 4) on S3D.}
    \label{tab:post-time}
    \resizebox{\linewidth}{!}{%
        \begin{tabular}{@{}lcccccccc@{}}
            \rotatebox{90}{}  & 
            \rotatebox{90}{CR}  &
            \rotatebox{90}{1. I/O} & \rotatebox{90}{\begin{tabular}[c]{@{}l@{}}2. Comp\&\\ Decomp\end{tabular}} & 
            \rotatebox{90}{3. \begin{tabular}[c]{@{}l@{}} Sample\\ + Model\end{tabular}} & 
            \rotatebox{90}{4. \begin{tabular}[c]{@{}l@{}} Process\end{tabular}}    & 
            \rotatebox{90}{5. \begin{tabular}[c]{@{}l@{}} Ori\\ (c1+c2)\end{tabular}}    &
            \rotatebox{90}{6. \begin{tabular}[c]{@{}l@{}} Extra\\ (c3+c4)\end{tabular}}  &
            \rotatebox{90}{\textbf{\begin{tabular}[c]{@{}l@{}}Overhead\\ (c6/c5)\end{tabular}}} 
            \\ \midrule
            \multirow{3}{*}{\begin{tabular}[c]{@{}l@{}}ZFP \\ (OpenMP)\end{tabular}} 
                & Small & 0.77                                             & 1.403                                                      & 0.009                                                       & 0.050                                                & 2.175                                                    & 0.059                                                      & \textbf{0.027}                                                      \\
                                                                                     & Mid   & 0.79                                             & 1.081                                                      & 0.010                                                       & 0.051                                                & 1.876                                                    & 0.061                                                      & \textbf{0.033}                                                      \\
                                                                                     & Large & 0.80                                             & 0.948                                                      & 0.012                                                       & 0.049                                                & 1.749                                                    & 0.061                                                      & \textbf{0.035}                                                      \\ \midrule
            \multirow{3}{*}{\begin{tabular}[c]{@{}l@{}}SZ2 \\ (OpenMP)\end{tabular}} 
                & Small & 0.78                                             & 0.411                                                      & 0.010                                                       & 0.034                                                & 1.190                                                    & 0.044                                                      & \textbf{0.037}                                                      \\
                                                                                     & Mid   & 0.84                                             & 0.371                                                      & 0.008                                                       & 0.034                                                & 1.208                                                    & 0.042                                                      & \textbf{0.035}                                                      \\
                                                                                     & Large & 0.79                                             & 0.283                                                      & 0.007                                                       & 0.033                                                & 1.072                                                    & 0.039                                                      & \textbf{0.037}                                                      \\ \midrule
            \multirow{3}{*}{\begin{tabular}[c]{@{}l@{}}SZ2\\ (Serial)\end{tabular}}  
                & Small & 0.82                                             & 5.199                                                      & 0.031                                                       & 0.042                                                & 6.015                                                    & 0.073                                                      & \textbf{0.012}                                                      \\
                                                                                     & Mid   & 0.81                                             & 4.585                                                      & 0.028                                                       & 0.041                                                & 5.399                                                    & 0.069                                                      & \textbf{0.013}                                                      \\
                                                                                     & Large & 0.85                                             & 3.637                                                      & 0.021                                                       & 0.039                                                & 4.485                                                    & 0.061                                                      & \textbf{0.013}                                                      \\
            \bottomrule
        \end{tabular}

    }
\end{table}

Specifically, the original compression workflow (columns 1 and 2) includes reading the original file, compression and decompression, and writing the decompressed file.
Our post-processing involves sampling, (de)compressing the sampled data, modeling the optimal parameter before compression (column 3), and post-processing after decompression (column 4).
The efficiency of our approach is due to the high parallel efficiency of the B\'ezier curve and our effective implementation. As detailed in column 3 of TABLE~\ref{tab:post-time}, our sampling and modeling process incurs very low overhead for SZ2/ZFP with OpenMP. For serial SZ2, the sampling and modeling times are higher due to slower (de)compression speed, which, further minimizes our relative overhead. Moreover, our post-processing speed is notably fast, as shown in column 4. Note that the post-processing speed for ZFP is slower due to its smaller block size compared to SZ2, which increases processing intensity.

\section{Conclusion and Future Work}
\label{sec:conclusion}
This paper introduces a workflow for multi-resolution data compression, applicable to both uniform and AMR simulations. Initially, the workflow employs a compression-oriented ROI extraction approach to enable multi-resolution methods for uniform data.
We further propose adaptive padding and dynamic processing to improve the efficiency of three distinct compressors for multi-resolution data and improve the compression ratio of SOTA approaches by up to 3.3$\times$ under the same data quality. In addition, an advanced uncertainty visualization method is integrated to evaluate the compression impacts.
In the future, we aim to investigate how to effectively apply our workflow to sparse data, given that each individual level of multi-resolution data essentially constitutes sparse data. 
We will also study how our workflow can preserve application-specific post-analysis quality such as Halo-finder.
Additionally, we plan to explore post-processing curves beyond the Bézier curve and incorporate other visualization methods (e.g., volume rendering) to expand the scope of our uncertainty visualization for compression.

\section*{Acknowledgement}

\small

James Ahrens, Pascal Grosset, and Jesus Pulido are employees of Triad National Security, LLC, which operates Los Alamos National Laboratory under Contract No. 89233218CNA000001 with the U.S. Department of Energy (DOE) and National Nuclear Security Administration (NNSA), and their work on this project was funded by the U.S. DOE Office of Science (SC), Office of Advanced Scientific Computing Research (ASCR), under contracts DE-AC02-06CH11357 and DE-AC02-05CH11231 and the Exascale Computing Project (ECP), Project Number: 17-SC-20-SC, a collaborative effort of the DOE SC and NNSA. 
Tushar Athawale's funding was supported by the U.S. Department of Energy (DOE) RAPIDS-2 SciDAC project under contract number DE-AC0500OR22725. 
Daoce Wang and Dingwen Tao's work on this project was supported by the National Science Foundation (Grant Nos. 2312673, 2247080, and 2303064).
Daoce Wang was also supported by the Exascale Computing Project (ECP), Project Number: 17-SC-20-SC, a collaborative effort of the DOE SC and NNSA during his summer internship at Los Alamos National Laboratory.
Dingwen Tao was also supported by the National Natural Science Foundation of China (Grant Nos. 62032023 and T2125013), and the Innovation Funding of ICT, CAS (Grant No. E461050). 
This material is also based upon work supported by the CAMPA collaboration, a project of the U.S. Department of Energy, Office of Science, Office of Advanced Scientific Computing Research and Office of High Energy Physics, Scientific Discovery through Advanced Computing (SciDAC) program.

\newpage

\renewcommand*{\bibfont}{\small}
\printbibliography[]

\end{document}